\documentclass[sigconf]{acmart}
\usepackage{fdiaz}

\usepackage[noend]{algpseudocode}
\usepackage{subcaption}
\usepackage[]{algorithm2e}

\copyrightyear{2022}
\acmYear{2022}
\setcopyright{rightsretained}
\acmConference[SIGIR '22]{Proceedings of the 45th International ACM SIGIR Conference on Research and Development in Information Retrieval}{July 11--15, 2022}{Madrid, Spain}
\acmBooktitle{Proceedings of the 45th International ACM SIGIR Conference on Research and Development in Information Retrieval (SIGIR '22), July 11--15, 2022, Madrid, Spain}\acmDOI{10.1145/3477495.3532033}
\acmISBN{978-1-4503-8732-3/22/07}

\newcommand{\query}[0]{q}
\newcommand{\queries}[0]{\mathcal{Q}}

\newcommand{\ndocs}[0]{n}
\newcommand{\nrel}[0]{m}
\newcommand{\rel}[0]{\fdvec{y}}

\newcommand{\rlx}[0]{\pi}
\newcommand{\rly}[0]{\pi'}

\newcommand{\irlx}[0]{\tilde{\pi}}
\newcommand{\irly}[0]{\tilde{\pi}'}
\newcommand{\rlset}[0]{S_{\ndocs}}
\newcommand{\rltest}[0]{\tilde{S}_{\ndocs}}

\newcommand{\metric}[0]{\mu}
\newcommand{\pmetric}[0]{\nu}
\newcommand{\dmetric}[0]{\Delta\mu}
\newcommand{\ap}[0]{\text{AP}}

\newcommand{\pp}[0]{\text{RPP}}
\newcommand{\interleaving}[0]{\text{I}}
\newcommand{\stpp}[0]{\text{ST-RPP}}
\newcommand{\pptest}[0]{\pp_{\rltest}}

\newcommand{\user}[0]{u}
\newcommand{\users}[0]{\fdset{U}}

\newcommand{\esl}[0]{f}

\newcommand{\asl}[1]{\text{ASL}(#1)}

\newcommand{\rpx}[0]{\esl}
\newcommand{\rpy}[0]{\esl'}

\newcommand{\gradeThreshold}[0]{\lambda}
\newcommand{\gradeThresholds}[0]{\Lambda}
\newcommand{\subtopic}[0]{t}
\newcommand{\subtopics}[0]{\mathcal{T}}

\newcommand{\discount}[0]{\delta}

\newcommand{\egx}[0]{X}
\newcommand{\egy}[0]{Y}

\newcommand{\st}[0]{\fdvec{s}}

\begin{document}
\fancyhead{}

\title{Offline Retrieval Evaluation Without Evaluation Metrics}

\author{Fernando Diaz}
\affiliation{
  \institution{Canadian CIFAR AI Chair}
  \country{}
}
\affiliation{%
  \institution{Google}
  \city{Montr\'eal}
  \state{QC}
  \country{Canada}
}
\email{diazf@acm.org}

\author{Andres Ferraro}
\affiliation{%
  \institution{Mila - Quebec Artificial Intelligence Institute}
  \city{Montr\'eal}
  \state{QC}
  \country{Canada}
}
\email{andresferraro@acm.org}

\begin{abstract}
    Offline evaluation of information retrieval and recommendation has traditionally focused on distilling the quality of a ranking into a scalar metric such as average precision or normalized discounted cumulative gain.  We can use this metric to compare the performance of multiple systems for the same request.  Although evaluation metrics provide a convenient summary of system performance, they also collapse subtle differences across users into a single number and can carry assumptions about user behavior and utility not supported across retrieval scenarios.  We propose recall-paired preference (RPP), a metric-free evaluation method based on directly computing a preference between ranked lists.  RPP simulates multiple user subpopulations per query and compares systems across these pseudo-populations.  Our results across multiple search and recommendation tasks demonstrate that RPP substantially improves discriminative power while correlating well with existing metrics and being equally robust to incomplete data.  
\end{abstract}


\maketitle
\section{Introduction}
\label{sec:introduction}

\begin{figure}
    \centering
    {
        \begin{subfigure}[b]{\linewidth}
            \centering
            \includegraphics[width=2in]{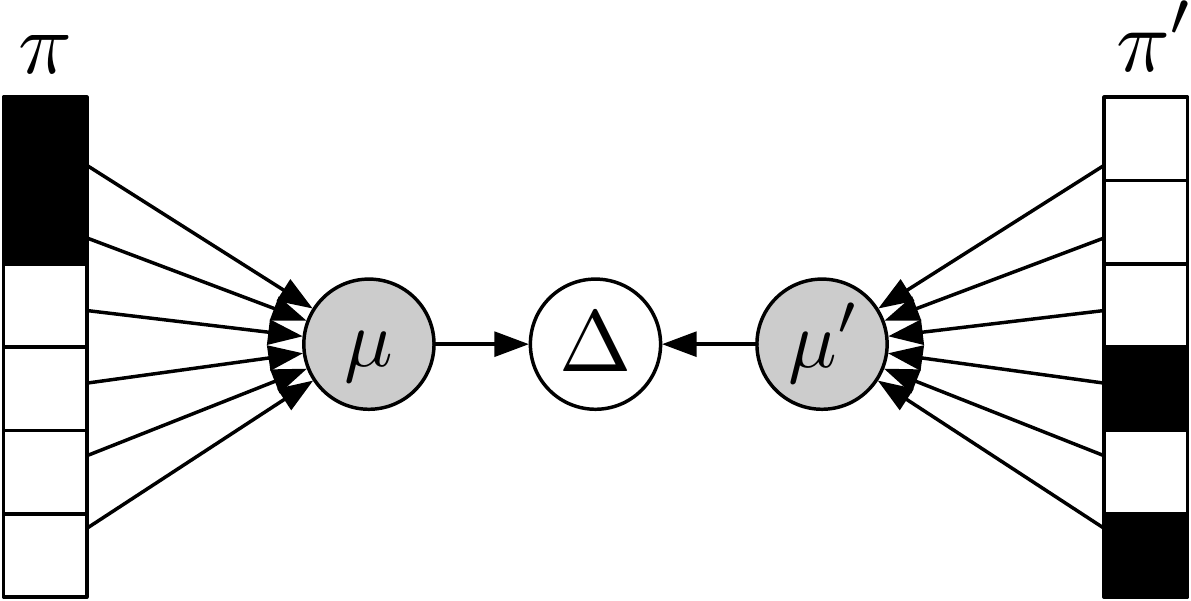}
            \caption{Metric Difference}\label{fig:metric-vs-preference:metric}
        \end{subfigure}
    }
    {
        \begin{subfigure}[b]{\linewidth}
            \centering
            \includegraphics[width=2in]{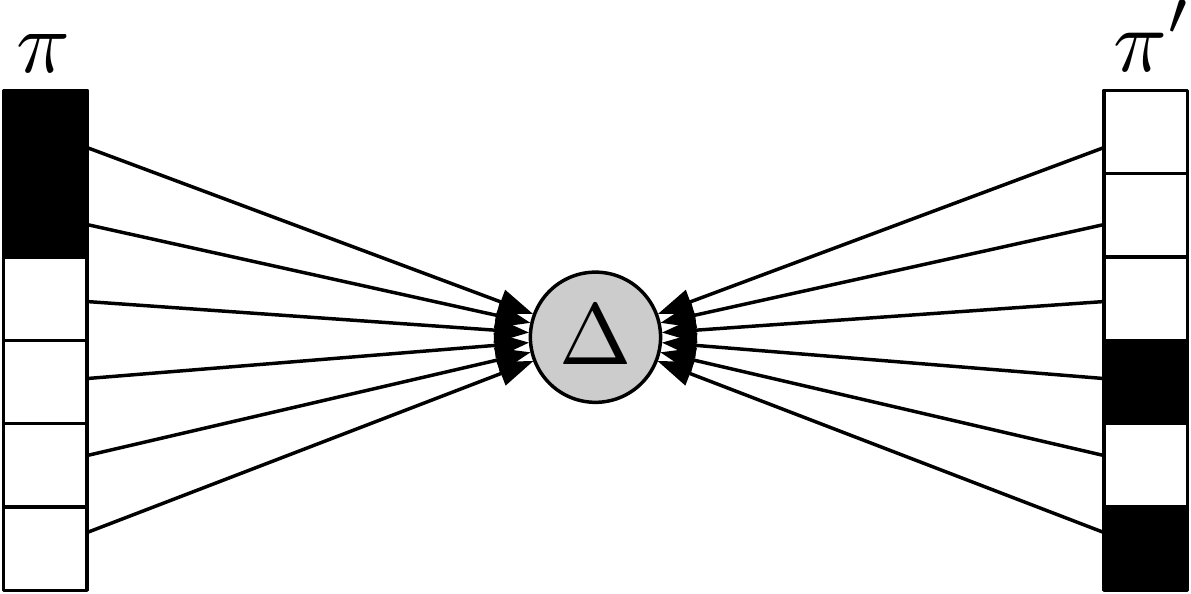}
            \caption{Direct Preference}\label{fig:metric-vs-preference:preference}
        \end{subfigure}
    }
    \caption{Metric Difference versus Direct Preference. System rankings $\rlx$ and $\rly$ are represented as boxes with shaded boxes indicating relevant item positions.  A traditional evaluation metric $\metric$ such as average precision projects two system rankings to scalar values; the scalar metric difference indicates preference.  Direct preference  compares ranked lists explicitly, bypassing metric computation.  Shaded nodes contrast the focus of research work between metrics and preferences. }\label{fig:metric-vs-preference}
\end{figure}

A fundamental step in the offline evaluation of search and recommendation systems is to determine whether a ranking from one system tends to be better than the ranking of a second system.  This often involves, given item-level relevance judgments, distilling each ranking into a scalar evaluation metric $\metric$, such as average precision (AP) or normalized discounted cumulative gain (NDCG).  We can then say that one system is preferred to another if its metric values tend to be higher.  We present a stylized version of this approach in Figure \ref{fig:metric-vs-preference:metric}.  

Deriving a system preference from a metric difference can be problematic for two reasons.  First, evaluation metrics, because they project a ranking onto a scalar value, can lose information about how two rankings differ.  Take, as an example, the popular reciprocal rank metric (RR).  Because RR only considers the rank position of the first relevant document, its value can be equal for two rankings that share the position of the first relevant document but differ dramatically at lower ranks.  Although most salient for RR, metrics with smooth discount functions such as AP and NDCG still can collapse different rankings into the same or very similar scalar value by virtue of their sharp discounts.  We refer to this as the problem of low \textit{label efficiency}. 
Second, although most evaluation metrics are meant to model the quality of a ranking for system users, they can suggest similarity between systems that actually behave quite differently for different user populations.  For example,  RR might be an appropriate model for known-item search, but it does not capture higher-recall behaviors like electronic discovery and systematic review \cite{zhang:high-recall}.  While metrics with smooth discounts can be interpreted as averaging performance across different possible user behaviors, they make very strong assumptions about the distribution of behaviors.  We refer to this as the problem of low \textit{robustness to user behavior}.  

We propose rank-paired preference (RPP), an evaluation method that addresses concerns both about label efficiency and robustness to user behavior.  For a fixed request, RPP directly computes a preference between systems by modeling how different user subpopulations might prefer one algorithm over another.  When aggregating these subpopulation preferences, each is weighted equally, rather than weighting users with lower recall requirements more heavily.  We contrast RPP with metric-based evaluation in Figure \ref{fig:metric-vs-preference:preference}.  By considering the contribution of lower-ranked relevant items, RPP more efficiently exploits available labels, resulting in higher sensitivity and discriminative power between systems compared to metric-based approaches.

We analyze RPP across a variety of search and recommendation tasks.  Specifically, we show that 
\begin{inlinelist}
    \item RPP is correlated with existing ranking metrics,
    \item RPP is equally robust to incomplete evaluation data compared to existing ranking metrics, and
    \item RPP has much higher discriminative power than existing ranking metrics.
\end{inlinelist}
In particular, RPP's higher discriminative power suggests that preference-based evaluation should be further explored for offline evaluation.

\section{Motivation}
\label{sec:motivation}
In order to motivate our work, consider a retrieval scenario with binary relevance.  Most ranked list evaluation metrics can be decomposed into a linear function of rank positions of the relevant items.  Given a system ranking $\rlx$, let $\rpx_i$ be the position of the $i$th relevant item.  We can define many metrics as,
\begin{align*}
    \metric(\rlx)&=\sum_{i}^\nrel \discount(\rpx_i)
\end{align*}
where $\nrel$ is the number of relevant items and $\discount$ is a rank discount function (e.g., $\discount_{\text{DCG}}(i)=\frac{1}{\log(i+1)}$, $\discount_{\text{RBP}}(i)=\gamma^{i-1}$).  In offline evaluation, we are interested in comparing this value to that of a second ranking $\rly$, using the difference in metric values to define preference.  We can expand this difference into a sum of differences over the $\nrel$ positions of the relevant items,
\begin{align*}
    \dmetric(\rlx,\rly)&=\sum_{i}^\nrel \discount(\rpx_i)-\sum_{i}^\nrel \discount(\rpy_i)\\
    &=\sum_{i}^\nrel \discount(\rpx_i)-\discount(\rpy_i)\\
    &=\sum_{i}^\nrel \dmetric_i(\rlx,\rly)
\end{align*}
This disaggregation by recall level lets us observe how the $i$th relevant item contributes to a change in $\dmetric$.  

\begin{figure}
    \centering
    \includegraphics[width=3.25in]{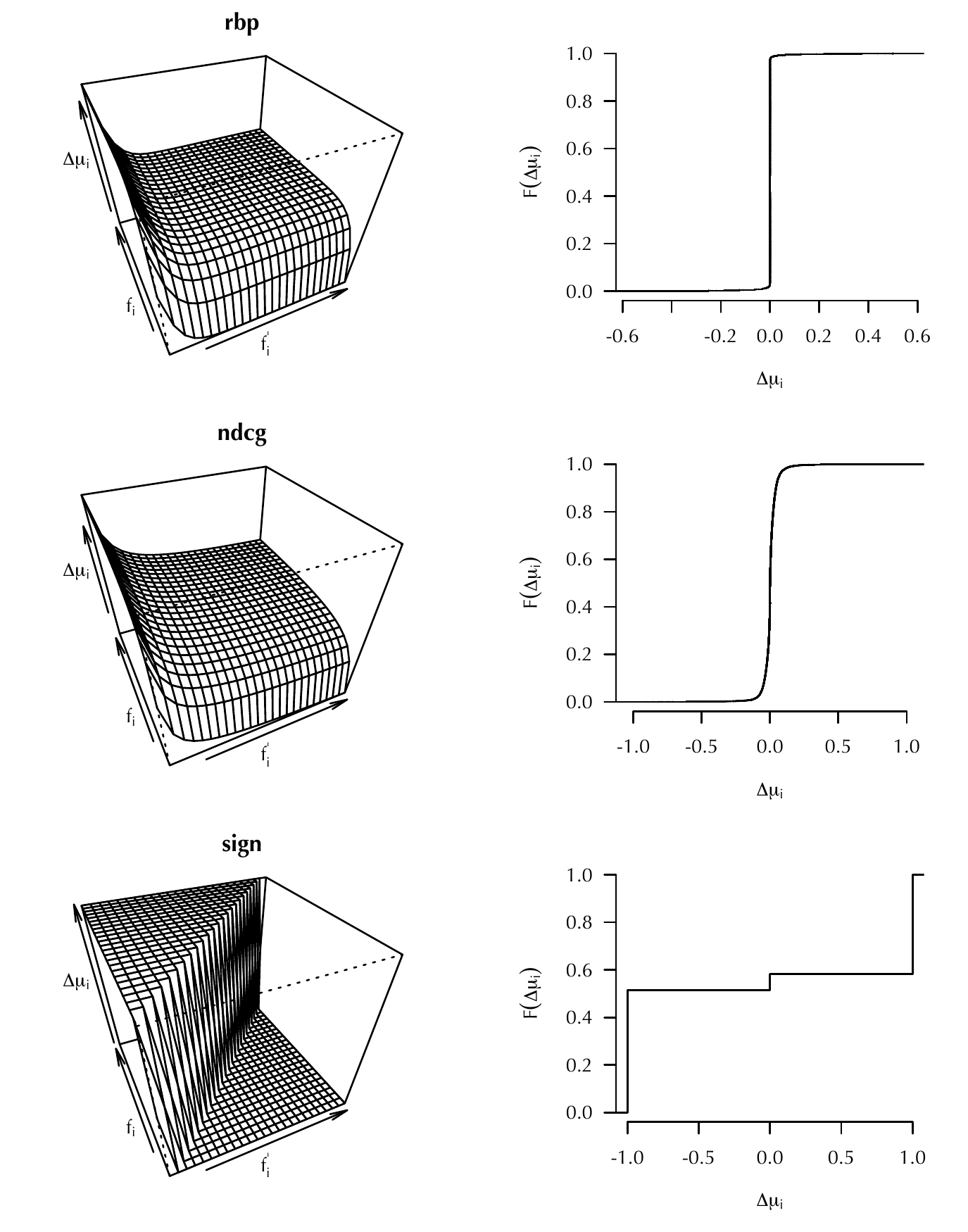}
    \caption{Surface of $\dmetric_i$ for comparing differences in the top 25 rank positions (left).  Empirical cumulative distribution function for differences observed in all runs submitted to the TREC 2019 Deep Learning document ranking task (right). }\label{fig:differences}
\end{figure}

In Figure \ref{fig:differences}, we examine the behavior of $\dmetric_i$ under different evaluation metrics.  In the left column, we show the relationship between $\dmetric_i$ and the position of the $i$th relevant item in the pair of  ranked lists being compared (i.e. $\rpx_i$ and $\rpy_i$); in other words, this is the analytic relationship between $\rpx_i$, $\rpy_i$, and $\dmetric_i$ for different evaluation metrics.  The first row considers rank-biased precision (RBP) with $\gamma=0.5$ \cite{moffat:rbp}; the second row, NDCG \cite{NDCG}; the last row,  $\dmetric_i=\text{sgn}(\discount(\rpx_i)-\discount(\rpy_i))$, reflecting the preference between rankings by a user seeking to find exactly $i$ relevant items with as little effort as possible.  Looking at the surface of $\dmetric_i$ for the RBP and NDCG metrics, we observe that, unless $\min(\rpx_i,\rpy_i)$ is small, the value of $\dmetric_i$ will be very small; as a result, the summation in $\dmetric$ will be dominated by changes in rank position amongst documents at the highest rank positions.  This means that the relative preference of systems by users interested in higher recall will be overshadowed by the preferences of users interested in fewer relevant items.

In the right column of Figure \ref{fig:differences}, we show the associated empirical cumulative distribution function of $\dmetric_i$ for all runs submitted to the TREC 2019 Deep Learning document ranking task \cite{craswell:trec-dl-2019}.  We can see that the distrbution of $\dmetric_i$ for RBP and NDCG is dominated by values close to zero.  Looking at the sign of these differences in the last row, we observe that, for roughly 93\% of the samples, $\rpx_i\neq\rpy_i$.

This analysis provides evidence that the discounting of lower-ranked relevant items massively diminishes their contribution to metric differences, resulting lower label efficiency and robustness to user behavior.  In the remainder of this paper, we  address this by presenting an alternative evaluation method that more equally incorporates preferences of users with  different recall requirements.

\section{Preliminaries}
\label{sec:preliminaries}
We are interested in comparing pairs of rankings for the same request from two different systems.  Our methods apply to both search and recommendation contexts.  As such, we will refer to queries and user profiles as \textit{requests} and documents as \textit{items}.  

Given a request $\query$ and a corpus of $\ndocs$ items, let $\rel\in\Re^{\ndocs}$ be a vector of item relevance grades, where $\rel_i$ is the relevance grade of item $i$.  We refer to $\rel$ as the relevance judgments for $\query$ and any grade greater than 0 as relevant.  Given a request $\query$, let $\rlx\in\rlset$ be a system's ranking of items in the corpus.  In cases where a system returns a ranking only of the top $k$ items, we assume a worst case ordering of the remaining items.  For example, if a top $k$ ranking $\egx$ omits five items with grades $\{1,1,2,4,5\}$ from the retrieval, then we treat these items as ranked in \textit{increasing} order of utility at the end of the total corpus ranking,
\begin{align*}
    \egx&=\underbrace{04300010300\ldots000}_{\text{top $k$}}\underbrace{0000\ldots0011245}_{\text{bottom $\ndocs-k$}}
\end{align*}  
Similarly, if a top $k$ ranking $\egy$ omits three items with grades $\{1,2,4\}$ from the retrieval,
\begin{align*}
    \egy&=\underbrace{30453001100\ldots000}_{\text{top $k$}}\underbrace{0000\ldots0000124}_{\text{bottom $\ndocs-k$}}
\end{align*}
This worst case assumption provides a conservative lower bound on system performance.

When provided with two rankings $\rlx$ and $\rly$ for the same request, we are interested in determining if we prefer $\rlx$ to $\rly$, indicated as $\rlx \succ \rly$.  \textit{Metric-based evaluation} leverages a function $\metric : \rlset \rightarrow \Re$ that computes the quality of a ranking defines a preference as,
\begin{align*}
    \metric(\rlx) > \metric(\rly) &\rightarrow \rlx \succ \rly\\
    \metric(\rlx) < \metric(\rly) &\rightarrow \rlx \prec \rly
\end{align*}
Our work considers \textit{preference-based evaluation}, an approach that directly defines a function $\Delta : \rlset\times\rlset\rightarrow\Re$ such that,
\begin{align*}
    \Delta(\rlx,\rly) > 0 &\rightarrow \rlx \succ \rly\\
    \Delta(\rlx,\rly) < 0 &\rightarrow \rlx \prec \rly
\end{align*}
In this section, we will review and extend three concepts from the existing literature: position-based metrics, pseudo-populations, and preference-based evaluation.  In Section \ref{sec:rpp}, we will synthesize these concepts into our new evaluation method, recall-paired preference.

\subsection{Position-Based Evaluation Metrics}
\label{sec:position-metrics}
Let the \textit{search length} $\esl_i(\rlx)$ be the position of the $i$th ranked relevant item in $\rlx$.\footnote{This is slightly different from \citeauthor{cooper:esl}'s definition of search length \cite{cooper:esl} which counts the number of non-relevant items above the $i$th relevant item, (i.e. $\esl_i(\rlx)-i$).}   In our example, $\esl_1(\egx)=2$, $\esl_2(\egx)=3$, $\esl_3(\egx)=7$, and so forth.    For clarity, we use $\rpx_i$ to refer to $\esl_i(\rlx)$ and $\rpy_i$ to refer to $\esl_i(\rly)$.  Position-based evaluation metrics adopt the principle of minimal effort,
\begin{quote}
    \emph{For a user interested in $i$ relevant items,}
    \begin{align*}
        \rpx_i<\rpy_i\rightarrow\rlx\succ\rly
    \end{align*}
\end{quote}

\citet{cooper:esl} describes different types of users who may be interested in different levels of recall $i$.  We refer to situations where a user is looking for one relevant item as \textit{precision-oriented}.  Historically, the reciprocal of the rank of the first relevant item (i.e. $\frac{1}{\esl_1}$) has been used for such tasks.  We call $\esl_i$ the \textit{initial search length} (ISL) and it is related to \citeauthor{cooper:esl}'s `Type 1 search length.' 

We refer to situations where a user is looking for all of the relevant items as \textit{recall-oriented}.  Assuming $\nrel$ relevant items, we refer to  $\esl_\nrel$, the position of the last relevant item, as the \textit{total search length} (TSL). This is related to \citeauthor{cooper:esl}'s `Type 3 inclusive search length.' 

In situations where we are uncertain if the user is precision- or recall-oriented, we can compute the expectation over all possible recall orientations.  We can express the \textit{average search length} (ASL) as,\footnote{Two other metrics, bpref \cite{buckley:evaluation-with-incomplete-information} and atomized search length \cite{alex:asl}, compute an expectation over relevant items, but, like \citeauthor{cooper:esl}, use the number of preceding nonrelevant items.  By linearity of expectation, we observe that $\mathbb{E}_i\left[\rpx_i-i\right] =\mathbb{E}_i\left[\rpx_i\right]-\mathbb{E}_i\left[i\right]$, which \citet{rocchio:thesis} refers as the \textit{recall error}.  For a fixed query, this value is rank-equivalent to $\mathbb{E}_i\left[\rpx_i\right]=\asl{\rlx}$.}
\begin{align}
    \asl{\rlx}&=\mathbb{E}_i\left[\rpx_i\right]\label{eq:asl}
\end{align}
\citet{rocchio:thesis} refers to this as the \textit{average rank} metric.

More recently, ASL has been used in the context of unbiased learning to rank \cite{joachims:unbiased-ltr-w-biased-fb}.

\subsection{Pseudo-Populations}
\label{sec:pseudopopulation}
In the previous section, we described metrics that operate under the assumption that a user is interested in precisely $i$ relevant items.  In order to build on this work, we turn to how we can model different possible subpopulations of users who desire different numbers of relevant items.

Similar to \citeauthor{cooper:esl}, \citet{robertson:ap-user-model} suggested that a user may have a specific recall requirement when interacting with an information access system.  In other words, for a  request with $\nrel$ relevant items, a user may be in one of $\nrel$ recall conditions. We refer to $\users_i$ as the \textit{pseudo-population} interested in $i$ relevant items and  $p(i)=p(\user\in\users_i)$ is the probability of a user seeking exactly $i$ relevant items.  \citeauthor{robertson:ap-user-model} observed that, if $p(i)=\frac{1}{\nrel}$, then AP can be interpreted as the expected precision for users from these pseudo-populations,
\begin{align*}
    \ap(\rlx) &= \mathbb{E}_i\left[\frac{i}{\rpx_i}\right]
\end{align*}

where $\frac{i}{\rpx_i}$ is the precision for users with a recall requirement of $i$.  \citet{sakai:user-models} extended this to arbitrary metrics, defining Normalized Cumulative Precision (NCP) as,
\begin{align*}
    \mathbb{E}_i\left[\metric_i(\rlx)\right] &= \sum_{i=1}^{\nrel} p(i)\metric_i(\rlx)
\end{align*}
where $\metric_i(\rlx)$ is a \textit{partial recall metric} based on the recall level $i$.  Example partial recall metrics include  $\metric_i(\rlx)=\frac{i}{\rpx_i}$ (i.e. precision at $\rpx_i$, as used in AP) and $\metric_i(\rlx)=\rpx_i$ (as used in ASL).  

The distribution $p(i)$ gives us a place where we can explicitly encode any information we have about user recall requirements.  For example, when we do not know the distribution of recall requirements of users or if we want robust performance across a variety of possible recall requirements, a uniform distribution over all recall levels (i.e. $p(i)=\frac{1}{m}$) might be appropriate.  

In  situations where we have information about the true distribution of recall requirements or want to emphasize performance for a specific user behavior, we can adopt a non-uniform distribution for $p(i)$.  For example,  `position bias' can be reflected in a definition of $p(i)$ that monotonically decreases with $i$.  In contrast with most existing models, this distribution is over recall levels, as opposed to rank positions.  

We can also consider pseudo-populations generated from other available information.  For example, when we have multiple possible relevance grades, we can consider a pseudo-population of users satisfied by an item if its grade is above some threshold \cite{robertson:graded-ap}.  

The utility for the pseudo-population of users interested in items with at least grade $\gradeThreshold$ is,
\begin{align*}
    \mathbb{E}_i\left[\metric_i(\rlx)|\gradeThreshold\right] &= \sum_{i=1}^{\nrel_{\gradeThreshold}} p(i|\gradeThreshold)\metric_{i,\gradeThreshold}(\rlx)
\end{align*}
where $\nrel_{\gradeThreshold}$ is the number of items with a grade of at least $\gradeThreshold$, $p(i|\gradeThreshold)$ is the probability of a user in this population seeks exactly $i$ relevant items, and $\metric_{i,\gradeThreshold}(\rlx)$ is the binary partial recall metric assuming only items with grade greater than $\gradeThreshold$ are relevant.

We will use $\gradeThreshold=1$ to refer to users who consider relevance as binary.  

In situations where items are associated with attributes such as genres or per-request subtopics, we can define pseudo-populations based on these categories \cite{agrawal:diversity}.  
As a result, the utility for the pseudo-population of users interested in category $\subtopic$ is,
\begin{align*}
    \mathbb{E}_i\left[\metric_i(\rlx)|\subtopic\right] &= \sum_{i=1}^{\nrel_{\subtopic}} p(i|\subtopic)\metric_{i,\subtopic}(\rlx)
\end{align*}
where $p(i|\subtopic)$ defines the probability that a user interested in subtopic $\subtopic$ seeks $i$ relevant items and $\metric_{i,\subtopic}(\rlx)$ is the binary partial recall metric assuming only items from subtopic $\subtopic$ are relevant.

\subsection{Preference-Based Evaluation}
\label{sec:preference}
Most existing approaches---including those in Sections \ref{sec:position-metrics} and \ref{sec:pseudopopulation}---collapse the performance of a system into a single scalar number.  An alternative to comparing metrics is to compare rankings directly.  

Preference-based evaluation\footnote{We note that preference-based evaluation differs from evaluating with item preferences, which  often still collapses rankings into a single scalar number \cite{carterette:preference-evaluation,clarke:top-k-preferences}.} assigns, for a pair of rankings, a preference between them.  Traditionally, we elicit this preference from human judges in an interface that presents two rankings alongside each other  \cite{thomas:sbs,kim:preference}.  \citet{sanderson:preferences} demonstrated that this approach correlated well with metric-based evaluation across a variety of retrieval scenarios.

In the context of online experimentation, interleaving combines pairs of rankings and computes a preference between them based on user clicks \cite{joachims:interleaving}. Given a request from a user, two rankings $\rlx$ and $\rly$ are randomly interleaved so as to simulate a choice experiment for the user.  The user then inspects the ranking, clicking on relevant items.  We say that the ranking $\rlx$ is preferred to $\rly$ if it retrieves more clicked items at a rank cutoff $k$, a value based on the position of the last-clicked item.  Because of randomness in both user behavior (e.g. their recall requirement) and the interleaving process itself, we can model $k$ as a random variable.  As a result, for a fixed query, we can define the interleaving preference as,
\begin{align}
    \interleaving(\rlx,\rly) &= \mathbb{E}_k\left[\text{sgn}(\pmetric_k(\rlx)-\pmetric_k(\rly))\right] \label{eq:interleaving}
\end{align}
where $\pmetric_k(\rlx)$ is a \textit{partial precision metric} based on the rank position $k$; we contrast this with partial recall metrics, which are based on recall level.  Example partial precision metrics include $\pmetric_k(\rlx)=\frac{|\{\rpx_i|\rpx_i\leq k\}|}{k}$ (i.e. `precision at $k$', as used in interleaving).  In online evaluation, we compute the interleaving metric by empirically estimating Equation \ref{eq:interleaving} from sampled user requests and clicks on interleaved rankings.  
\citet{chapelle:interleaving} demonstrated the sensitivity of interleaving experiments across a variety of online search scenarios.

\section{Recall-Paired Preference}
\label{sec:rpp}
We are now ready to combine the concepts from Section \ref{sec:preliminaries} into a new preference-based evaluation method.

\begin{figure}
    \includegraphics[height=1.5in]{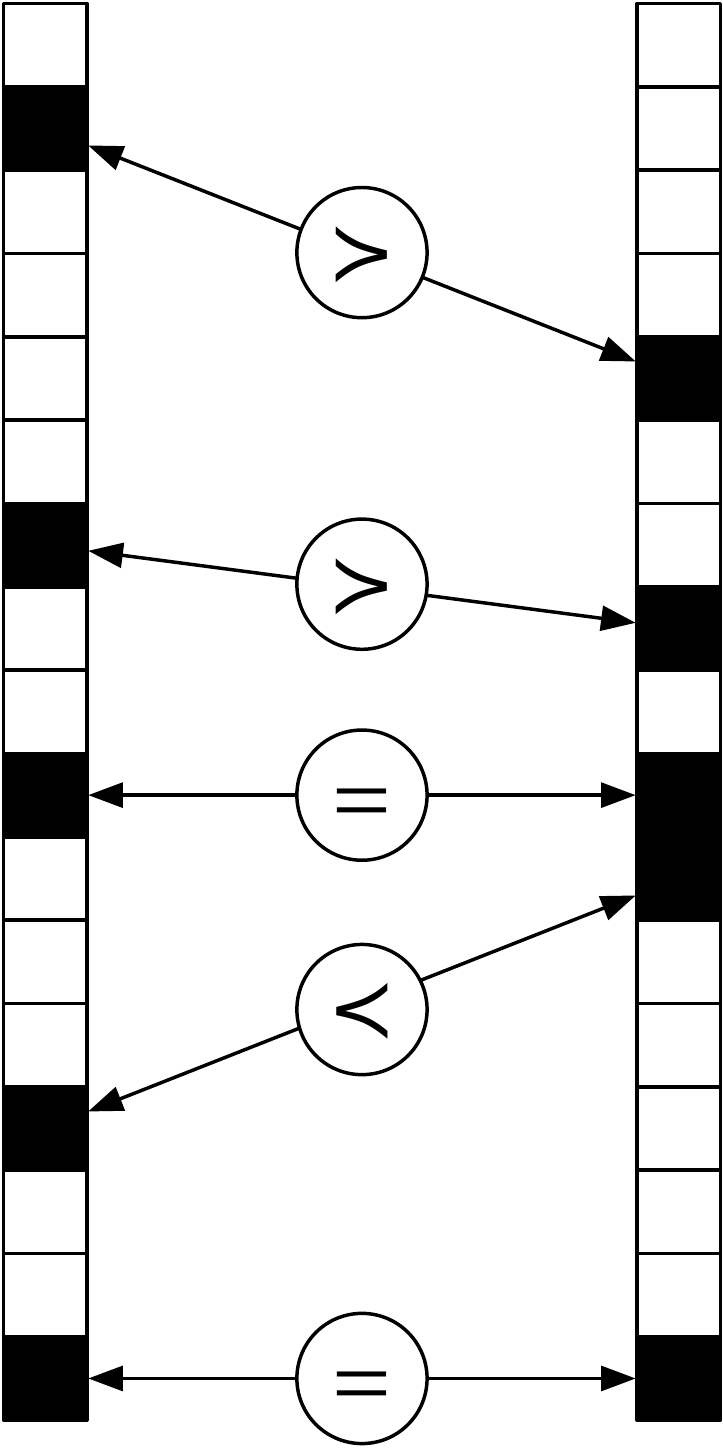}
    \caption{Recall-paired preference.  A user interested in $i$ relevant items will prefer the ranking that lets them satisfy their need with minimal effort.  }\label{fig:paired-preferences}
\end{figure}

We begin by describing conceptually how we compare two rankings.  Consider our example rankings $\egx$ and $\egy$ introduced in Section \ref{sec:preliminaries}.  First, we sample a user $\user$ based on our distribution over pseudo-populations.  By appealing to the  principle of minimal effort (Section \ref{sec:position-metrics}), we can infer which ranking $\user$ prefers.  For example, if we sample a user from $\users_1^{\gradeThreshold=1}$, then $\rpx_1(\egx) = 2 > 1 = \rpx_1(\egy)$ and, since we prefer higher ranks,  $\egy \succ \egx$.  If we sample a user from $\users_1^{\gradeThreshold=4}$, then, using $\rpx_{1,\gradeThreshold=4}(\rlx)$ to represent the position of the first ranked item with relevance grade at least 4, $\rpx_{1,\gradeThreshold=4}(\egx) = 2 < 3 = \rpy_{1,\gradeThreshold=4}(\egx)$  and $\egx\succ\egy$.  We can repeatedly sample users, incrementing an accumulator by 1 if $\egx \succ \egy$ and decrementing  by 1 if $\egx\prec \egy$. If the value of accumulator is positive, we say that $\egx \succ \egy$; if it is negative, then $\egx\prec \egy$.  This is equivalent to computing the expected preference across the $\nrel$ paired positions of relevant items (Figure \ref{fig:paired-preferences}). Because we pair items according to equivalent recall levels, we refer to this metric as \textit{recall-paired preference} (RPP).

More formally, for binary relevance and no subtopics, we define RPP between two rankings as the expected value of the preference,
\begin{align}
    \pp(\rlx,\rly) &= \mathbb{E}_i\left[\text{sgn}(\rpy_i-\rpx_i)\right]\\
    &=\sum_{i=1}^{\nrel} p(i)\times\text{sgn}(\rpy_i-\rpx_i)\label{eq:rpp}
\end{align}
where $p(i)$ is the probability of a user seeking exactly $i$ relevant items.    RPP takes a value in $[-1,1]$ where positive values indicate stronger preference for $\rlx$, negative values a preference for $\rly$, and zero indicating indifference.  Moreover, RPP is a preference, so $\pp(\rlx,\rly)=-\pp(\rly,\rlx)$.  

In practice, when we refer to RPP, we will use the graded version,
\begin{align}
    \pp(\rlx,\rly) &= \sum_{\gradeThreshold\in\gradeThresholds}\sum_{i=1}^{\nrel} p(i,\gradeThreshold)\times\text{sgn}(\rpy_{i,\gradeThreshold}-\rpx_{i,\gradeThreshold})
\end{align}
where $\gradeThresholds$ is the set of all possible grades for this request and $\rpx_{i,\gradeThreshold}$ is the rank position of the $i$th relevant item with grade of at least $\gradeThreshold$.  In the binary relevance case, this reduces to Equation \ref{eq:rpp}.

The subtopic-aware version of RPP can be similarly defined,
\begin{align}
    \stpp(\rlx,\rly) &= \sum_{\subtopic\in\subtopics}\sum_{i=1}^{\nrel} p(i,\subtopic)\times\text{sgn}(\rpy_{i,\subtopic}-\rpx_{i,\subtopic})
\end{align}
where $\subtopics$ is the set of all possible subtopics for this request and $\rpx_{i,\subtopic}$ is the rank position of the $i$th relevant item with subtopic $\subtopic$.

\subsection{Comparison to Existing Metrics}
In this section, we compare RPP to the methods presented in Section \ref{sec:preliminaries}.

\begin{figure}
    \centering
    {
        \renewcommand{\arraystretch}{1.5}

    \begin{subtable}[b]{\linewidth}
        \centering
        \caption{Disaggregated recall-paired metrics}\label{tab:disaggregated-metrics}
        {\footnotesize
        \begin{tabular}{lc}
            metric & $\dmetric_i$\\
            \hline
            RBP \cite{moffat:rbp} & $\gamma^{\rpx_i} - \gamma^{\rpy_i}$\\
            AP & $i\left(\frac{1}{\rpx_i} - \frac{1}{\rpy_i}\right)$\\
            NDCG  \cite{NDCG}& $\frac{1}{\log_2(\rpx_i+1)} - \frac{1}{\log_2(\rpy_i+1)}$\\
            RR  & $\frac{1}{\rpx_i} - \frac{1}{\rpy_i}, (i=1)$ \\
            ISL \cite{cooper:esl} & $\rpy_i - \rpx_i, (i=1)$\\
            TSL \cite{cooper:esl} & $\rpy_i - \rpx_i, (i=\nrel)$\\
            ASL, bpref \cite{rocchio:thesis,buckley:evaluation-with-incomplete-information,alex:asl}& $\rpy_i - \rpx_i$\\
            RPP & $\text{sgn}(\rpy_i - \rpx_i)$\\
            \hline
        \end{tabular}
        }
    \end{subtable}
    }

    \begin{subfigure}[b]{\linewidth}
        \centering
        \includegraphics[width=3.25in]{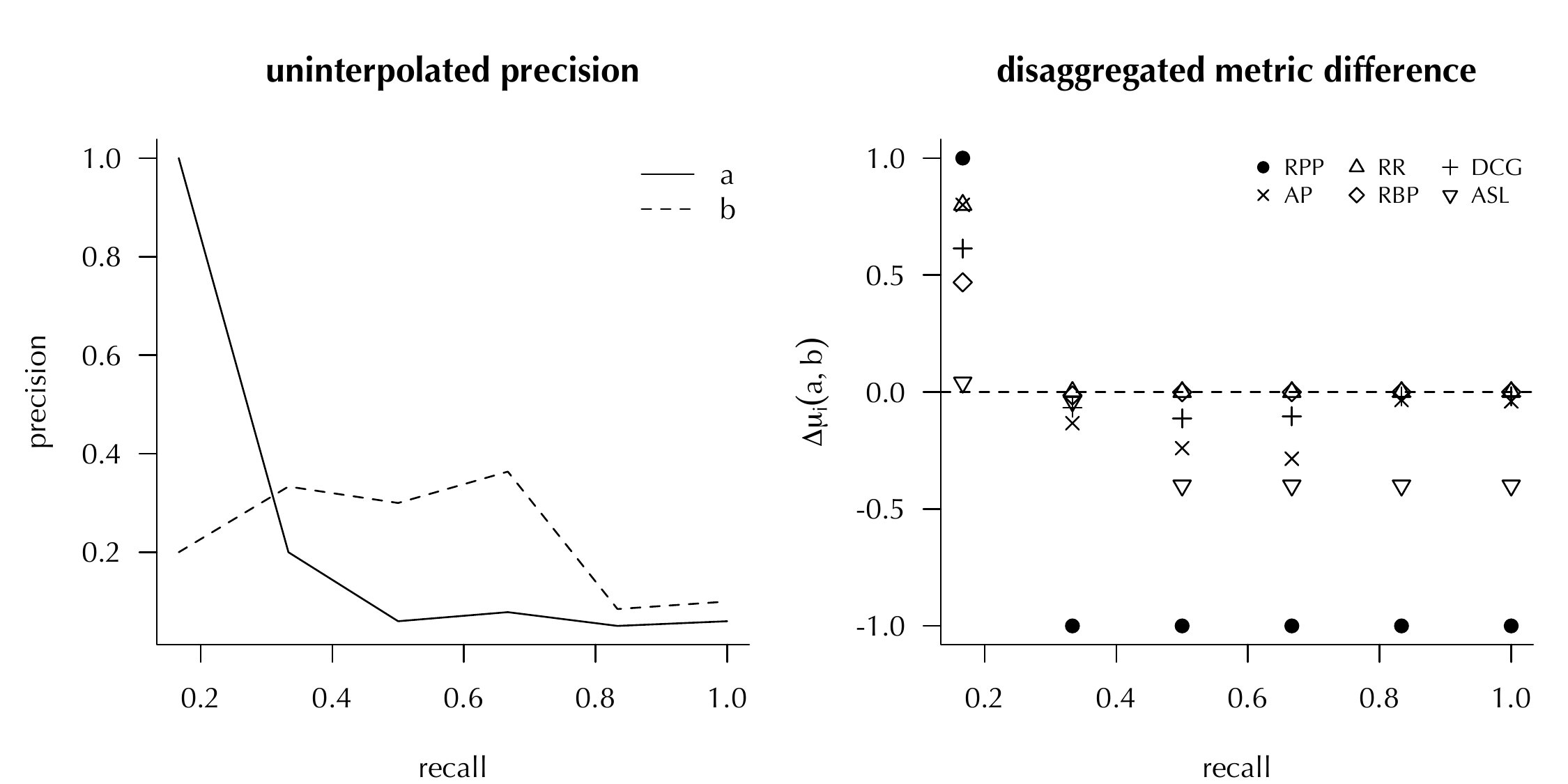}
        \caption{Metric differences for two systems' rankings of one request with six relevant items. }\label{fig:delta-metrics}
    \end{subfigure}
    \caption{Disaggregated metric behavior.}
\end{figure}

To begin, we can compare RPP with metric-based evaluation by analyzing how the disaggregated metric $\dmetric_i$ (Section \ref{sec:motivation}) changes as a function of $i$.  Figure \ref{tab:disaggregated-metrics} contains the disaggregated values for several standard evaluation metrics.  These expressions encode the relative weight allocated to different pseudo-populations $\users_i$.  Standard evaluation metrics such as RBP, AP, DCG, and RR all observe the largest contribution to metric differences at the highest rank positions.  Even AP, which modulates the difference in inverse positions with a multiplicative recall level factor $i$ is dominated by diminishing differences at low ranks.  This confirms our claim of poor label efficiency (since relevant items at lower rank positions can contribute less)
and poor robustness to user behavior (since the performance difference for users interested in higher recall levels is negligible).  RR, ISL, and TSL also exhibit poor label efficiency and robustness to user behavior since, by design, they exclude all but a single difference.  Meanwhile, ASL will tend to have the opposite effect of emphasizing differences at lower ranks since, at these higher recall levels, $\rpx_i$ and $\rpy_i$ are likely to be separated by many more rank positions than earlier recall levels.   In contrast, for RPP, the disaggregated magnitude is fixed across recall levels, resulting in both label efficiency (since all relevant items contribute equally regardless of rank position) and robustness to user behavior (since the performance differences for users at all recall levels contribute equally).

To illustrate the implication of these position biases, we can see how $\dmetric_i$ changes for two hypothetical rankings, $a$ and $b$.  In the left-hand plot of Figure \ref{fig:delta-metrics}, we depict the uninterpolated precision-recall curves for $a$ and $b$.    Whereas AP approximates the area under the  uninterpolated precision-recall curve, RPP can be interpreted as a sign test at sampled recall levels, similar to approaches taken for ROC curves \cite{braun:roc-sign}.

In the right-hand plot of Figure \ref{fig:delta-metrics}, we present $\dmetric_i(a,b)$ as a function of $i$.  Notice that almost every traditional metric, including `recall-oriented' metrics like AP, allocate most of the mass to the top position.  Conversely, ASL allocates more weight to later values of $i$.  RPP, on the other hand, treats all recall levels equally.

Although position-based metrics like ASL more evenly allocate weight across recall levels, \citet{magdy:pres} note that the range of this metric can be quite large, sensitive to outliers, and difficult to reason with.  Unretrieved items, often assumed to be ranked at the bottom of a ranking of the full corpus (Section \ref{sec:preliminaries}), can exacerbate the variance in the metric, even for moderately sized corpora.  In contrast, RPP, because it only considers relative positions, can be computed without an exact corpus size.

In comparison to interleaving, while Equations \ref{eq:interleaving} and \ref{eq:rpp} appear  similar, there are a few important differences.  First, the event space for interleaving is the set of all rank positions rather than the set of all recall levels.  This subtle difference means that interleaving emphasizes the number of relevant items collected at a rank position rather than the effort taken to collect the same number of relevant items.  Moreover, in online interleaving, because $p(k)$ is derived from behavioral data, it will be skewed toward higher rank positions (i.e. users introduce position bias) and preferences at lower positions will be overshadowed by those in higher positions.

\section{Experiments}
\label{sec:methods}

Our main thesis is that RPP, by more uniformly measuring performance across recall levels,  more efficiently uses relevance labels in evaluation compared to existing retrieval metrics. As such, experiments are centered around three questions,
\begin{inlinelist}
    \item how does RPP correlate with existing metrics?
    \item how robust is RPP to incomplete data?
    \item how effective is RPP at discriminating between runs?
\end{inlinelist}
In order to answer these questions, we use a variety of information access scenarios covering both search and recommendation tasks.  
\subsection{Data}
We present details of the data used in our experiments in Table \ref{tab:data}.  We include runs submitted to multiple TREC tracks, including the Deep Learning Document Ranking (2019, 2020), Deep Learning Passage Ranking (2019, 2020), Common Core (2017, 2018), Web (2009-2014), and Robust (2004).  We downloaded all data from NIST, including runs and relevance judgments.  Web track data includes subtopic judgments.  

Additionally, we used a variety of recommendation systems runs prepared by \citet{valcarce:recsys-ranking-metrics-journal} for the MovieLens 1M, LibraryThing, and Beer Advocate datasets.\footnote{\url{https://github.com/dvalcarce/evalMetrics}}  Consistent with their work, we converted graded judgments to binary labels by considering any rating below 4 as nonrelevant and otherwise relevant.

\begin{table}
    \caption{Datasets used in experiments.}\label{tab:data}
{\small
    \begin{tabular}{lcccc}
        &   requests    &   runs    &   rel/request & subtopics/request\\
        \hline
core (2017)	&	50  &  75   &   180.04  &   0\\
core (2018)	&	50  &  72   &   78.96  &   0\\
deep-docs (2019)	&	43  &  38 & 153.42  &   0\\
deep-docs (2020)	&	45  &  64 & 39.27  &   0\\
deep-pass (2019)	&	43  &  37 & 95.40  &   0\\
deep-pass (2020)	&	54  &  59 & 66.78  &   0\\
web (2009)	&	50  &  48 & 129.98  &   4.98\\
web (2010)	&	48  &  32 & 187.63  &   4.17\\
web (2011)	&	50  &  62 & 167.56  &   3.36\\
web (2012)	&	50  &  48 & 187.36  &   3.90\\
web (2013)	&	50  &  61 & 182.42  &   3.18\\
web (2014)	&	50  &  30 & 212.58  &   3.12\\
robust	&	249 &  110 &    69.93  &   0\\
ml-1M	&	6005    &  21 & 18.87  &   0\\
libraryThing &   7227    &   21  &   13.15   &   0\\
beerAdvocate &   17564    &   21  &   13.66   &   0\\
    \end{tabular}
}
\end{table}

\subsection{Methods}
In order to measure the similarity between RPP and metric-based approaches, we measured the Kendall's $\tau$ correlation between system ordering by RPP with system ordering by baseline metrics (Section \ref{sec:experiments:baseline-metrics}).  We describe how to compute an ordering of systems from RPP in Section \ref{sec:methods:aggregating-rpp}.

We evaluated the robustness to incomplete data  under two conditions.  Our first experiment tests how well a metric with fewer \textit{judged requests} can order systems compared to the same metric with the complete set of judged requests.  This simulates the scenario where we have a paucity of requests but, for those requests, we have ample labeled items.  Our second experiment tests how well a metric with fewer \textit{judgments per request} can order systems compared to the same metric with the complete set of judgments.  This simulates the scenario where we have ample requests but sparse judgments for each request.

In order to evaluate the sensitivity of a metric, we adopt \citeauthor{sakai:metrics}'s  method of computing discriminative power \cite{sakai:metrics}.  For a single data set (row in Table \ref{tab:data}), we compute the RPP or metric differences for all pairs of runs over all requests.  We then measure what fraction of system pairs achieve a $p$-value lower than 0.05 for each metric.  In order to compute $p$-values, we use two methods: a Student's $t$-test with Bonferonni correction and Tukey's honestly significant difference (HSD) test.  We adopt the randomized HSD as proposed by   \citet{carterette:multiple-testing}.

\subsection{RPP Variants}
For binary relevance with no subtopics, we consider several definitions of $p(i)$.  In addition to $p(i)=\frac{1}{\nrel}$, we include versions that consider non-uniform, top-heavy distributions of pseudo-populations,
\begin{align*}
    p_{\text{DCG}}(i)&\propto \frac{1}{\log_2(i+1)} &     p_{\text{inverse}}(i)&\propto \frac{1}{i}
\end{align*}
which reflect the rank importance for NDCG and RR.  Note that these discounts are a function of recall level, rather than rank position.

When we adopt graded RPP for evaluation, we assume independence between recall requirements and grade, $p(i,\gradeThreshold)=p(i)p(\gradeThreshold)$, and define $p(\gradeThreshold)$ for $\gradeThreshold\in\gradeThresholds$ as,
\begin{align*}
    p(\gradeThreshold)&\propto|\{i|\rel_i\geq\gradeThreshold\}|
\end{align*}

When conducting subtopic evaluation, we again assume independence between recall requirements and subtopic, $p(i,\subtopic)=p(i)p(\subtopic)$, with $p(\subtopic)$ defined as,
\begin{align*}
        p(\subtopic)&\propto|\{i|\rel_i>0\land\subtopic\in\st_i\}|
\end{align*}
where $\st_i\subseteq\subtopics$ indicates the subtopics associated with item $i$.  In addition to these $|\subtopics|$ pseudo-populations, we consider a background interest $\subtopic^*$ pseudo-population satisfied by \textit{any} subtopic (i.e. standard relevance, $p(\subtopic^*)\propto|\{i|\rel_i>0\}|$).

\subsection{Aggregating RPP}
\label{sec:methods:aggregating-rpp}
RPP gives us a preference between a pair of rankings  for the same request but we are often interested in generating an ordering of more than two systems.  Given a set of runs $\rltest$ for the same request, we can compute the win rate for a ranking $\rlx\in\rltest$ as,
\begin{align}
    \pptest(\rlx)&=\sum_{\rly\in\rltest}\pp(\rlx,\rly)
\end{align}
We can then use a preference aggregation scheme to order systems for a set of queries $\queries$.  In experiments where we need an ordering of systems for a set of requests,  we adopt Markov chain aggregation, due to its effectiveness across a variety of domains \cite{dwork:rank-aggregation}.  Note that $\pptest(\rlx)$ reflects the relative position of $\rlx$ within $\rltest$ and may vary across different sets of runs.

\subsection{Baseline Metrics}
\label{sec:experiments:baseline-metrics}
As baseline metrics, we used AP, NDCG, and RR with no rank cutoff as implemented in NIST \texttt{trec\_eval}.\footnote{\url{https://github.com/usnistgov/trec_eval}}  We implemented ASL by moving unranked items to the end of the corpus, using a corpus size equal to the number of unique items in union of all rankings and the relevant items.  For subtopic metrics, we use intent-aware mean average precision (MAP-IA) with no rank cutoff and intent-aware expected reciprocal rank (ERR-IA) and subtopic recall (strec), both with a rank cutoff of 20, as adopted for Web tracks and implemented in \texttt{ndeval}.\footnote{\url{https://github.com/trec-web/trec-web-2014}}

In order to compare with interleaving, we developed \textit{offline interleaving} (OI) based on a simulated user.  \citet{carterette:user-models-effectiveness} observed that many existing metric definitions implicitly include a model of user behavior.  As such, given relevance information and a pair of system rankings to compare, we can simulate user interaction and compute an offline interleaving preference.  To do so, given two rankings $\rlx$ and $\rly$, we can generate the two possible interleaved rankings $\irlx$ and $\irly$. Then, we can use a browsing model and relevance information to simulate an online interleaving experiment and estimate Equation \ref{eq:interleaving}.

\section{Results}
\label{sec:results}

\subsection{Correlation with Existing Metrics}
\label{sec:results:metric-correlation}

\begin{figure}[t]
    \centering
    \includegraphics[width=1.65in]{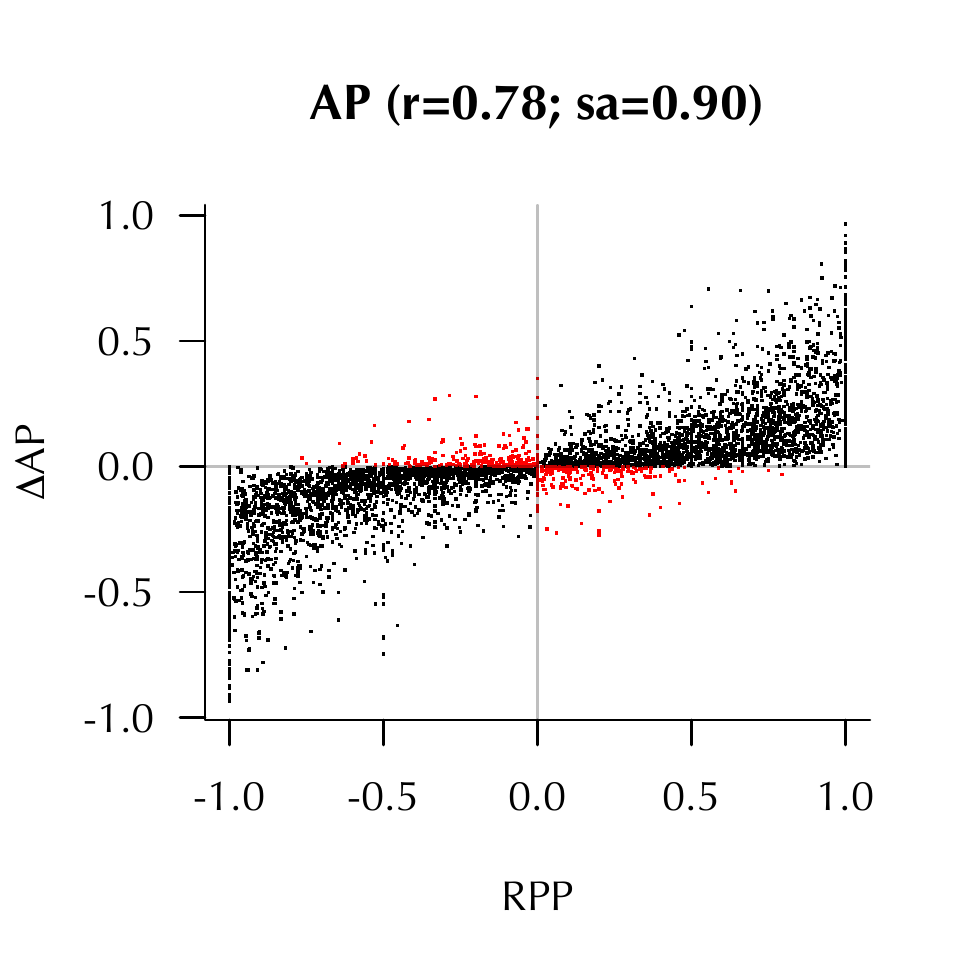}
    \includegraphics[width=1.65in]{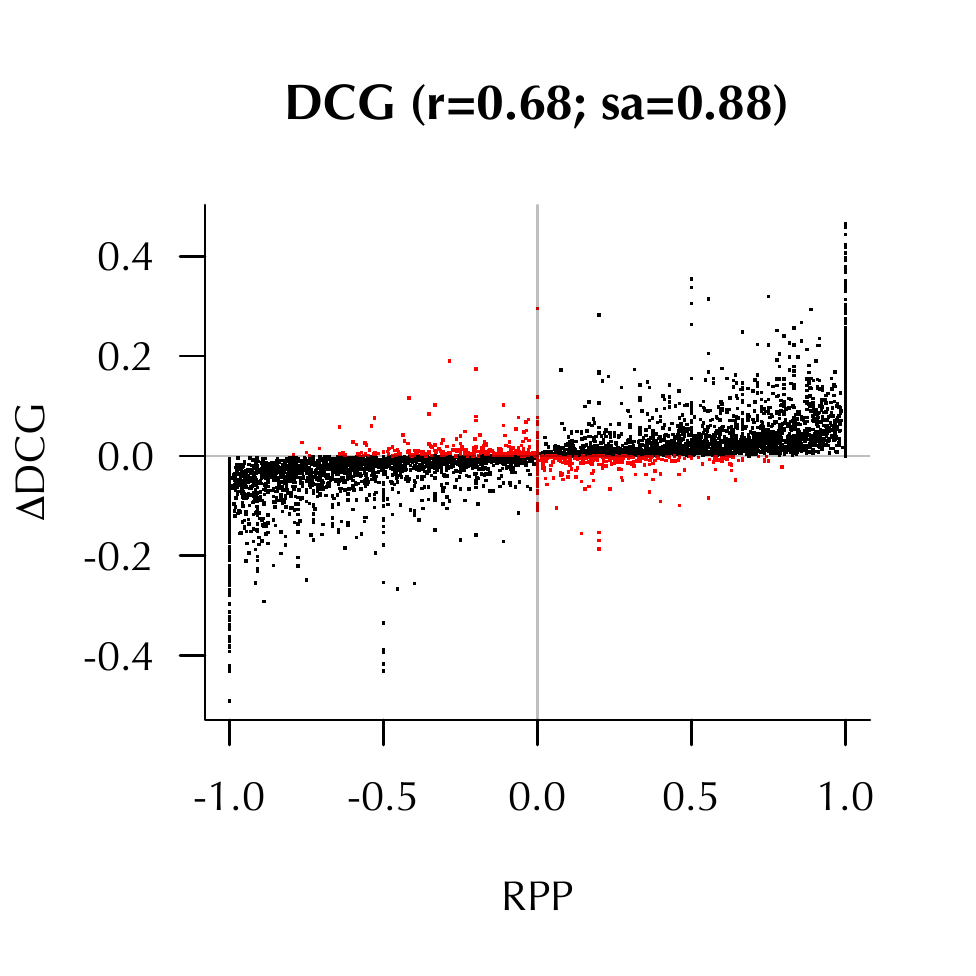}

    \includegraphics[width=1.65in]{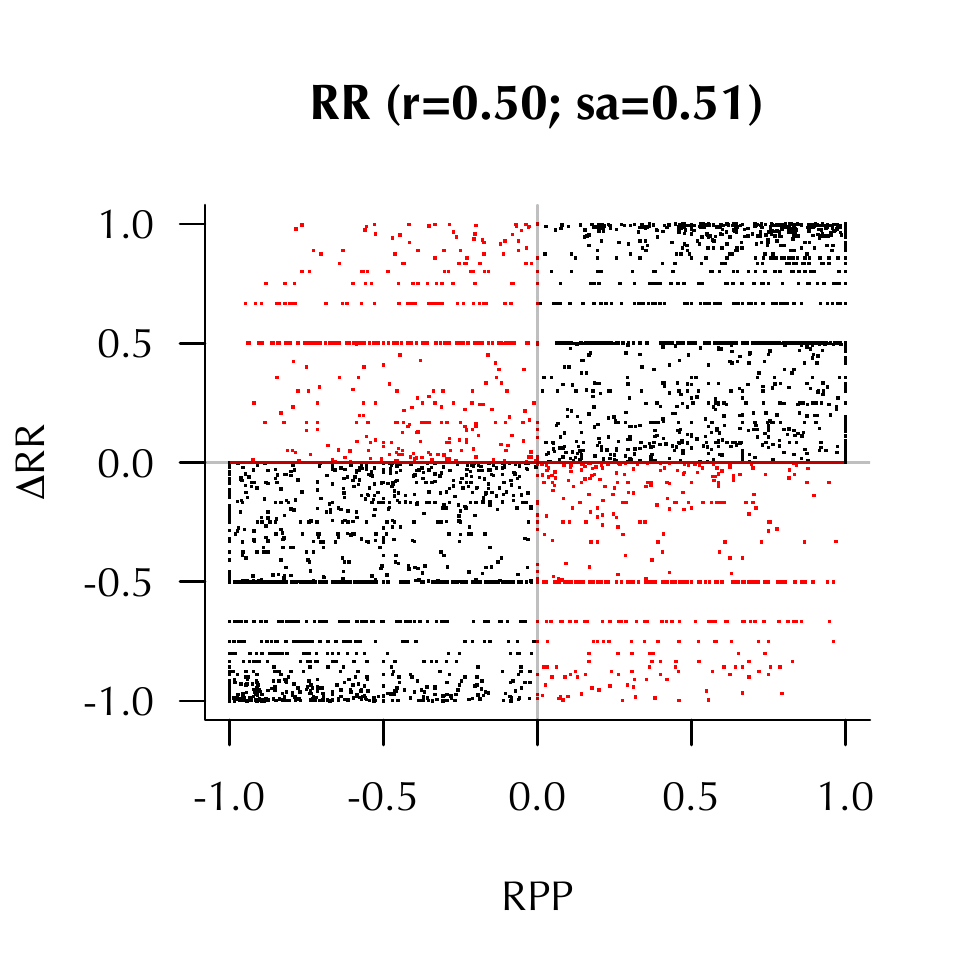}
    \includegraphics[width=1.65in]{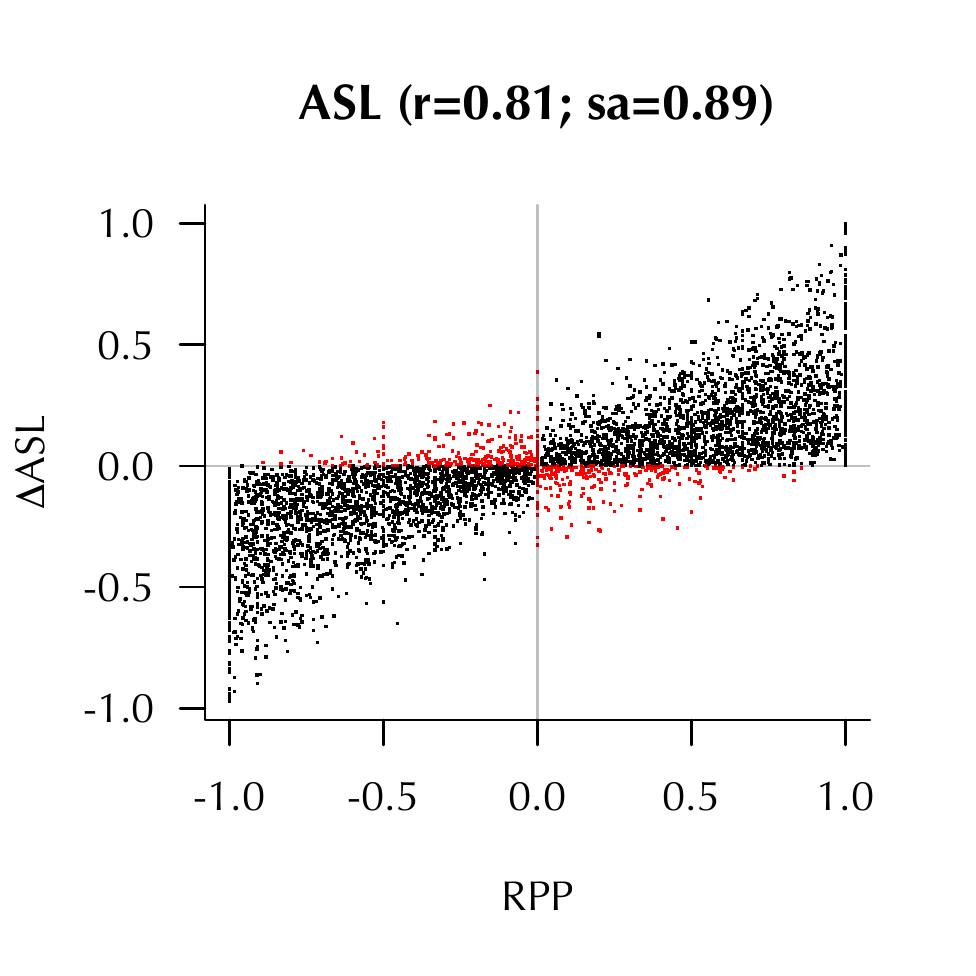}

    \includegraphics[width=1.65in]{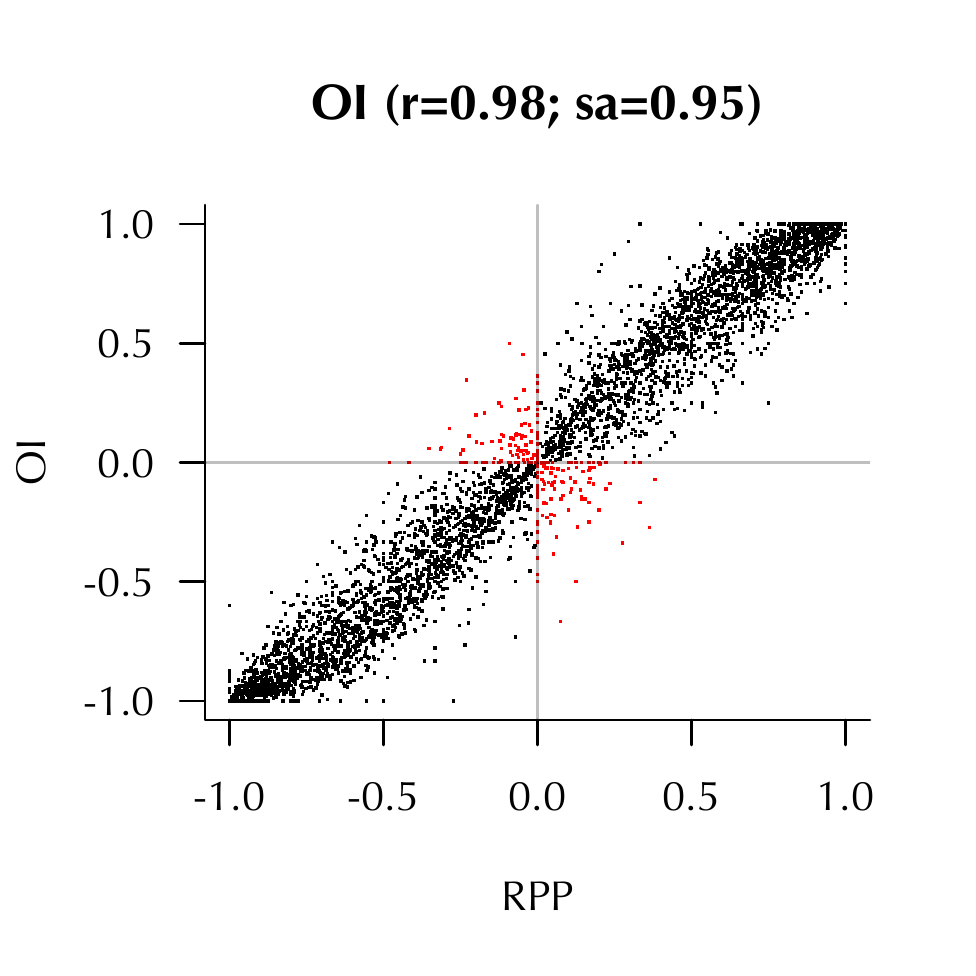}
    \includegraphics[width=1.65in]{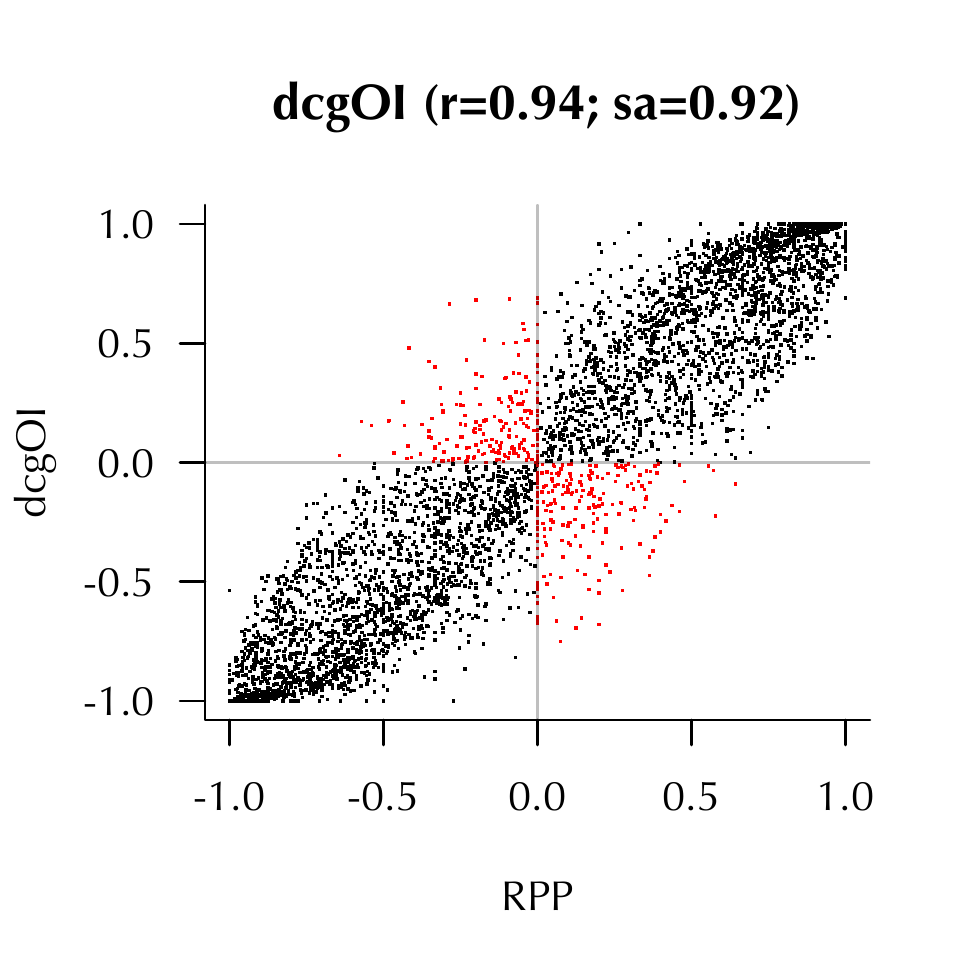}
    \caption{Query-level metrics differences for sampled runs from Robust.  Points in red indicate a difference in run ordering. Titles include the Pearson correlation (r) and fraction of points where RPP and the metric difference agree in sign (sa).}\label{fig:deltas}
\end{figure}

\begin{table}[t]
    \caption{Correlation with Existing Metrics.  Kendall's $\tau$ between rankings of runs for pairs of metrics averaged across all datasets. }\label{fig:metric-correlation}
    \centering
    {\small
        \begin{subtable}[b]{\linewidth}
            \caption{Single Topic Metrics}\label{fig:metric-correlation:single-topic}
            \centering
            \begin{tabular}{rccccc}
                \hline
                &   invRPP  &   RPP &   dcgRPP  &   AP  &   NDCG    \\
                \hline
                RR      &   0.61    &   0.45    &   0.49    &   0.49    &   0.50    \\
                invRPP   &   -   &   0.79    &   0.85    &   0.82    &   0.80    \\
                RPP    &   -   &   -   &   0.93    &   0.86    &   0.87    \\
                dcgRPP   &   -   &   -   &   -   &   0.88    &   0.88    \\
                AP   &   -   &   -   &   -   &   -    &   0.89    \\
                \hline
            \end{tabular}
        \end{subtable}
    }
    {\small
        \begin{subtable}[b]{\linewidth}
            \caption{Subtopic Metrics}\label{fig:metric-correlation:sub-topic}
            \centering
            \begin{tabular}{rccccc}
                \hline
                &   st-dcgRPP  &   MAP-IA &   st-invRPP  &   ERR-IA  &   strec    \\
                \hline
                st-RPP      &   0.88    &   0.73    &   0.70    &   0.26    &   0.30    \\
                st-dcgRPP   &   -   &   0.77    &   0.79    &   0.34    &   0.36    \\
                MAP-IA    &   -   &   -   &   0.74    &   0.39    &   0.36    \\
                st-invRPP   &   -   &   -   &   -   &   0.52    &   0.49    \\
                ERR-IA   &   -   &   -   &   -   &   -    &   0.66    \\
                \hline
            \end{tabular}
        \end{subtable}
    }
\end{table}

In order to get a sense of the relationship between baseline metric differences and RPP, we sampled pairs of runs for random queries in the Robust dataset.  We computed baseline metrics for each ranking and then plotted the metric differences against the RPP for the same pair of runs for the same query (Figure \ref{fig:deltas}).   Although the sign agreement of RPP with AP, NDCG, and ASL is close to 0.90, it drops to 0.50 for RR.  This result is consistent with the sign agreement between RR and AP (0.56), NDCG (0.58), and ASL (0.47).  

These results can be explained by two properties of RR.  First, because RR ignores recall levels higher than $1$, metrics that measure higher recall levels will incorporate information that can reverse the order of systems.  Second, the small number of unique RR values results in a number of ties between systems which are often resolved by metrics that consider more recall levels.

The sign \textit{disagreement} between RPP and AP, NDCG, and ASL tends to occur for small differences in performance, with metrics largely agreeing for dramatic differences in performance. This indicates that, even when the top ranked relevant items largely agree in classic metrics, there is enough disagreement at higher recall levels to differ from RPP.  

We implemented offline interleaving with two different user models, uniform and DCG-based.  We present the relationship between OI and RPP preference in the bottom row of Figure \ref{fig:deltas}.  We found that the sign agreement was higher between RPP and OI compared to RPP and baseline metrics. Moreover, the relationship between the preference magnitudes shows a strong linear correlated ($r=0.98, p<0.001$).  Comparing to OI with a DCG-style user model, the agreement and correlation degrade ($r=0.94, p<0.001$) but are still higher than baseline metrics.  

In addition to pairwise preference agreement, we were interested in the similarity between an ordering of runs induced from RPP preferences (Section \ref{sec:methods:aggregating-rpp}) and an ordering induced from baseline metrics.  To this end, we computed the Kendall's $\tau$ between the rankings of runs for each dataset.  We present the average correlation across these datasets in Table \ref{fig:metric-correlation:single-topic}.  We first notice that RPP with uniform position weighting (labeled RPP) correlates with AP and NDCG at a level close to how those metrics correlate with each other.  If we replace the uniform position weighting with a DCG-style non-uniform position weighting (dcgRPP), this correlation improves close to their correlation with each other.  As suggested by Figure \ref{fig:deltas}, the correlation between RR and RPP, AP, and NDCG is low.  Although correlation between dcgRPP with RR (0.49) is higher than RPP with RR (0.45), it remains comparable to that of RR with  AP and NDCG.  Using the reciprocal rank for the position discount (invRPP) improves this correlation further (0.61), suggesting that our pseudo-population modeling works as expected.  That said, we do not expect this metric to correlate perfectly with RR since invRPP considers items below the first relevant item.  

We also include correlations for subtopic metrics in Table \ref{fig:metric-correlation:sub-topic}.  We observe similar patterns to single topic metrics.  MAP-IA, a subtopic version of AP, correlates well with the subtopic versions of RPP and dcgRPP.  Top-heavy metrics ERR-IA and strec, on the other hand, correlate well with the subtopic version of invRPP, consistent with earlier results.  

Taken together, these results indicate that RPP metrics effectively capture a variety of aspects of baseline metrics but do not correlate perfectly, suggesting that they add information to evaluation.  Moreover, they demonstrate the ability to adapt RPP to different scenarios (e.g. position bias, novelty).

\subsection{Robustness to Incomplete Data}
\label{sec:results:robustness}
Because missing data is common in offline evaluation, we explored the behavior of RPP under two degradation schemes.  Due to space constraints, we provide representative results for news search (Robust), web search (2020 Deep Learning Passage Ranking), and recommendation (MovieLens 1M).

\begin{figure}
    \centering
    {
        \begin{subfigure}[b]{\linewidth}
            \centering
            \includegraphics[width=2.5in]{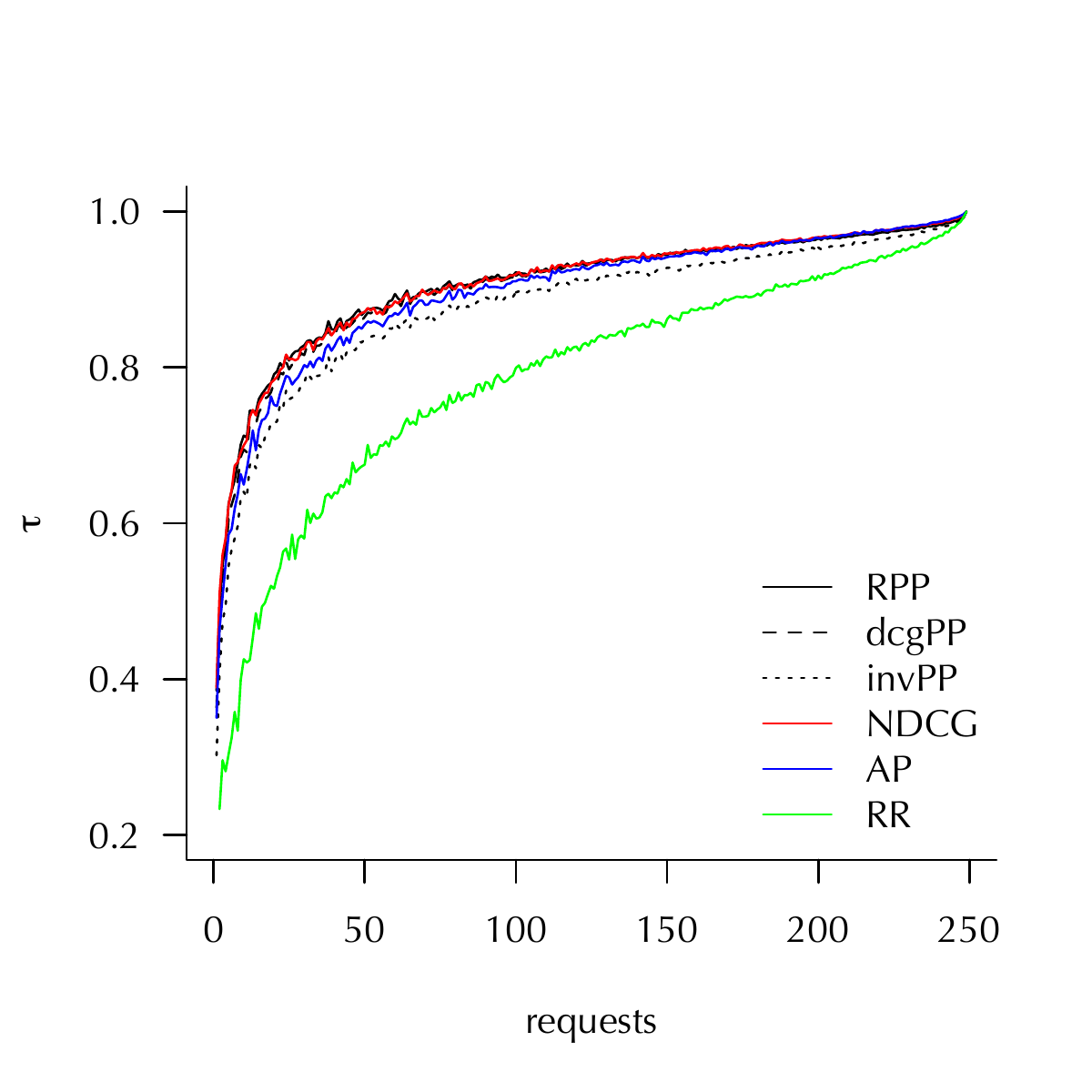}
            \caption{Robust}
        \end{subfigure}
    }
    {
        \begin{subfigure}[b]{\linewidth}
            \centering
            \includegraphics[width=2.5in]{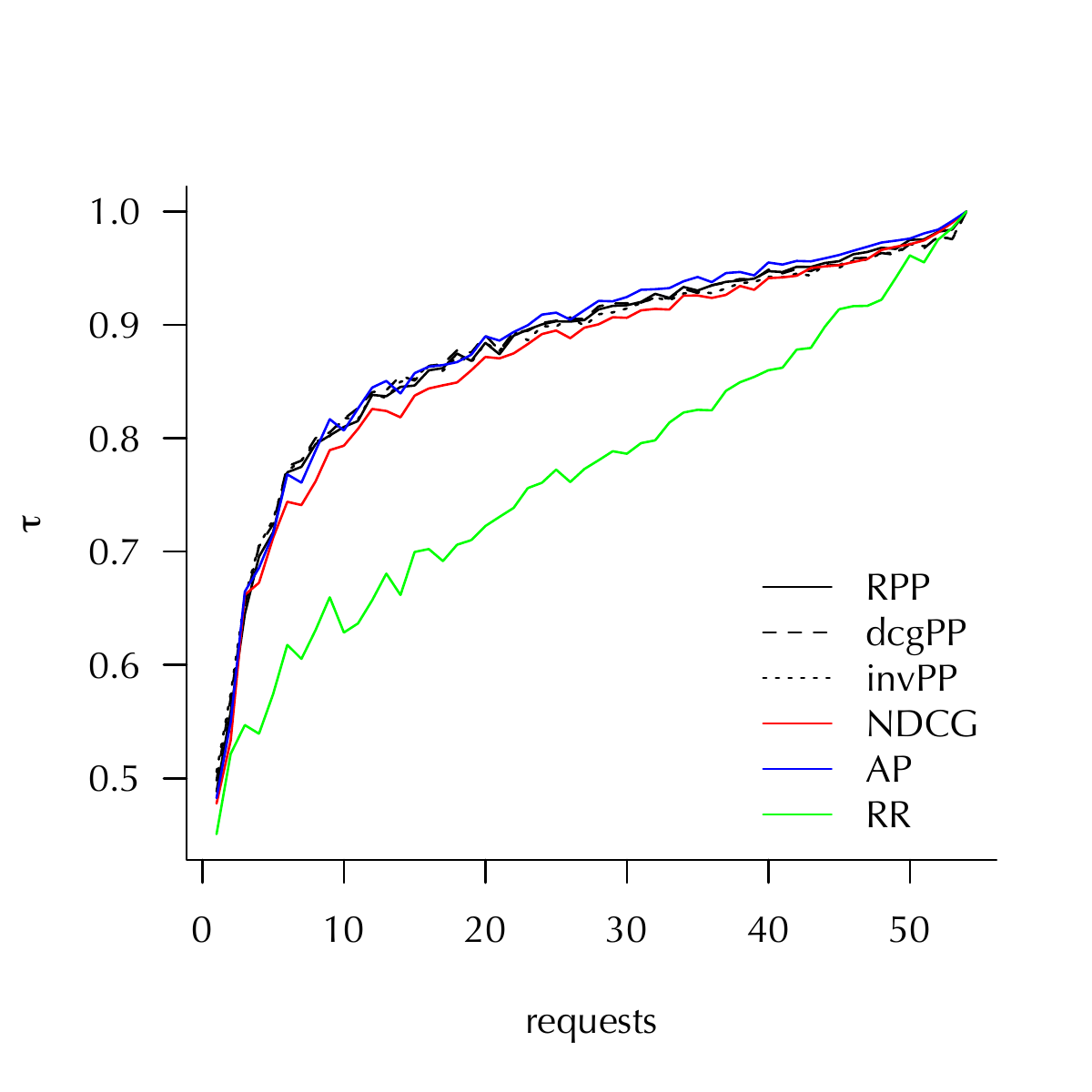}
            \caption{Deep Learning Passage Ranking (2020)}
        \end{subfigure}
    }
    {
        \begin{subfigure}[b]{\linewidth}
            \centering
            \includegraphics[width=2.5in]{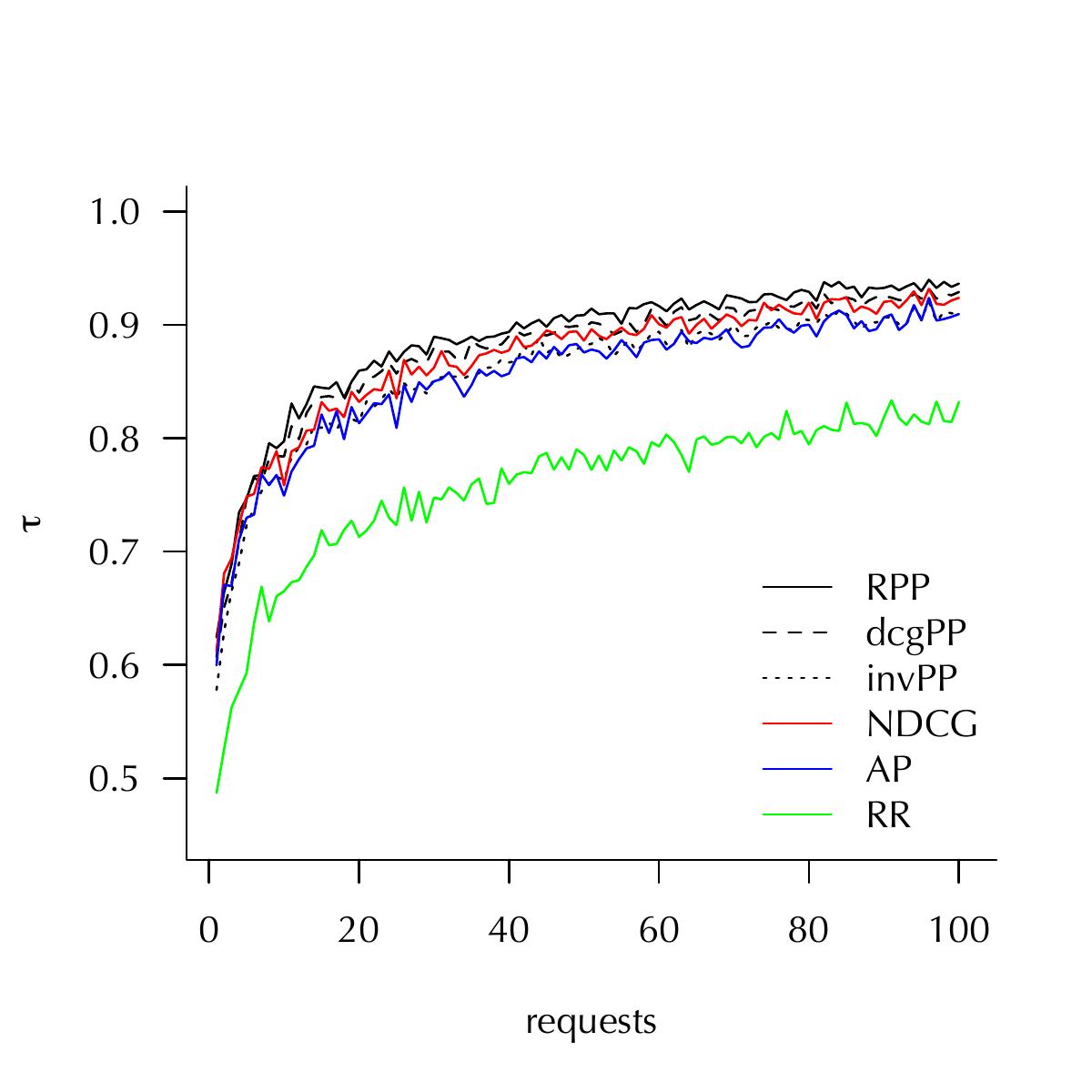}
            \caption{MovieLens 1M}
        \end{subfigure}
    }
    \caption{Kendall's $\tau$ of a system ranking given $k$ requests with a system ranking given all requests. Only the first 100 users are shown for ml-1M since metrics converge quickly after.  }\label{fig:dtau}
\end{figure}

We show the sensitivity of results when evaluating with fewer requests in  Figure \ref{fig:dtau}.  For each metric, we calculate the correlation between system rankings with missing requests and system rankings with all requests; an insensitive metric will have higher correlation with fewer requests.  Consistent with other work, across all datasets, RR correlation degrades the fastest, suggesting that removing a few requests can alter system ordering.  In general, the RPP family of metrics degrades as gracefully as or better than existing metrics, AP and NDCG.

\begin{figure}
    \centering
    {
        \begin{subfigure}[b]{\linewidth}
            \centering
            \includegraphics[width=2.5in]{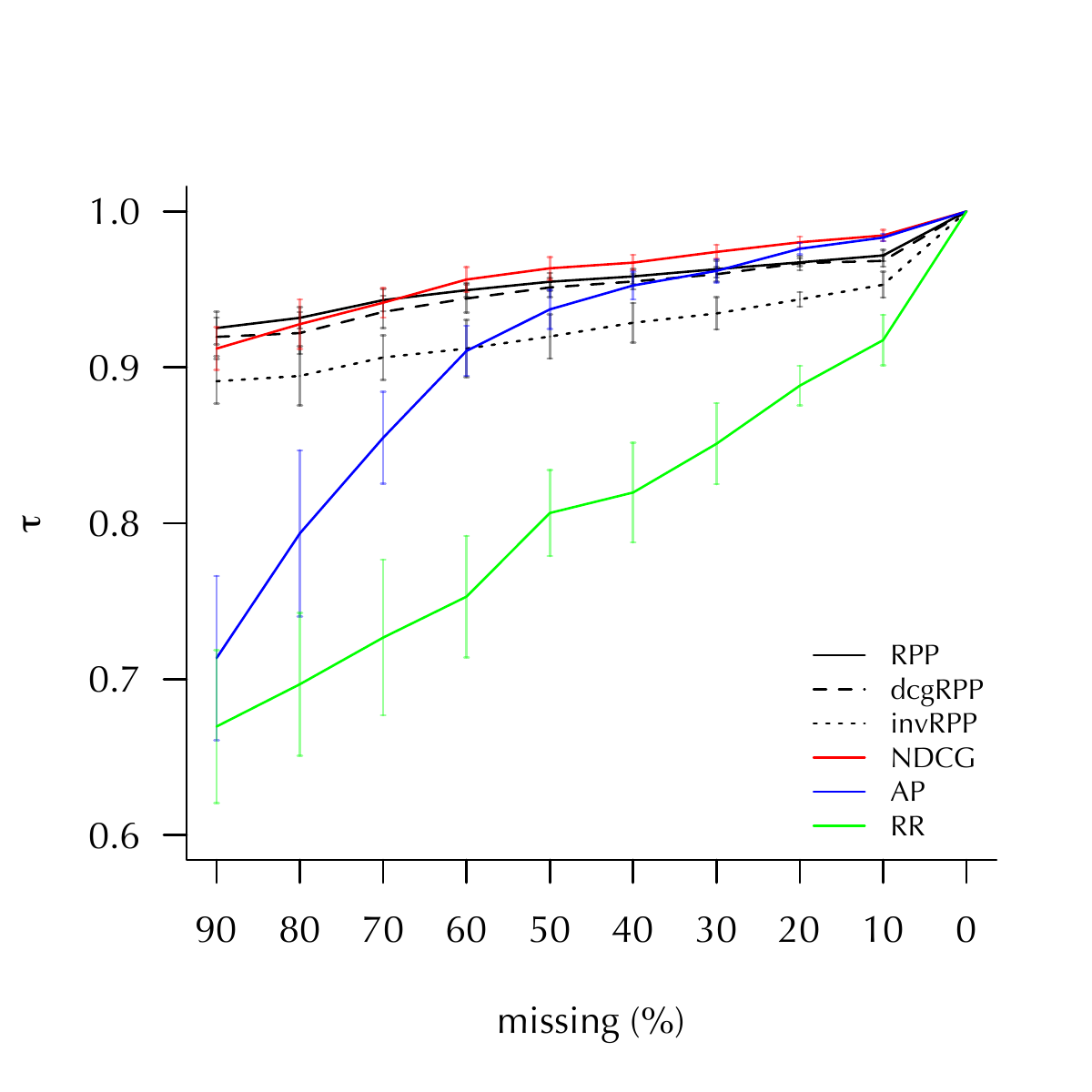}
            \caption{Robust}
        \end{subfigure}
    }
    {
        \begin{subfigure}[b]{\linewidth}
            \centering
            \includegraphics[width=2.5in]{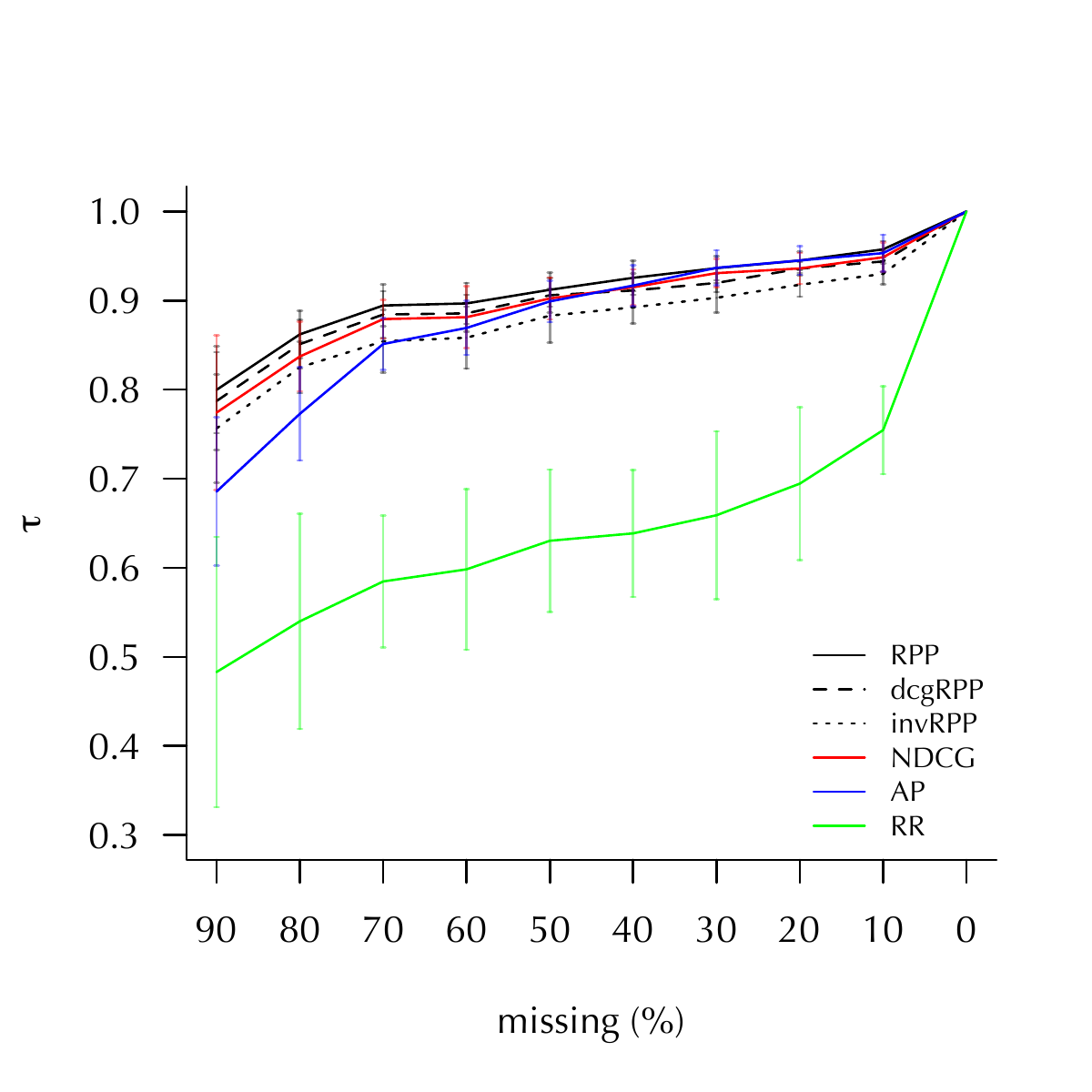}
            \caption{Deep Learning Passage Ranking (2020)}
        \end{subfigure}
    }
    {
        \begin{subfigure}[b]{\linewidth}
            \centering
            \includegraphics[width=2.5in]{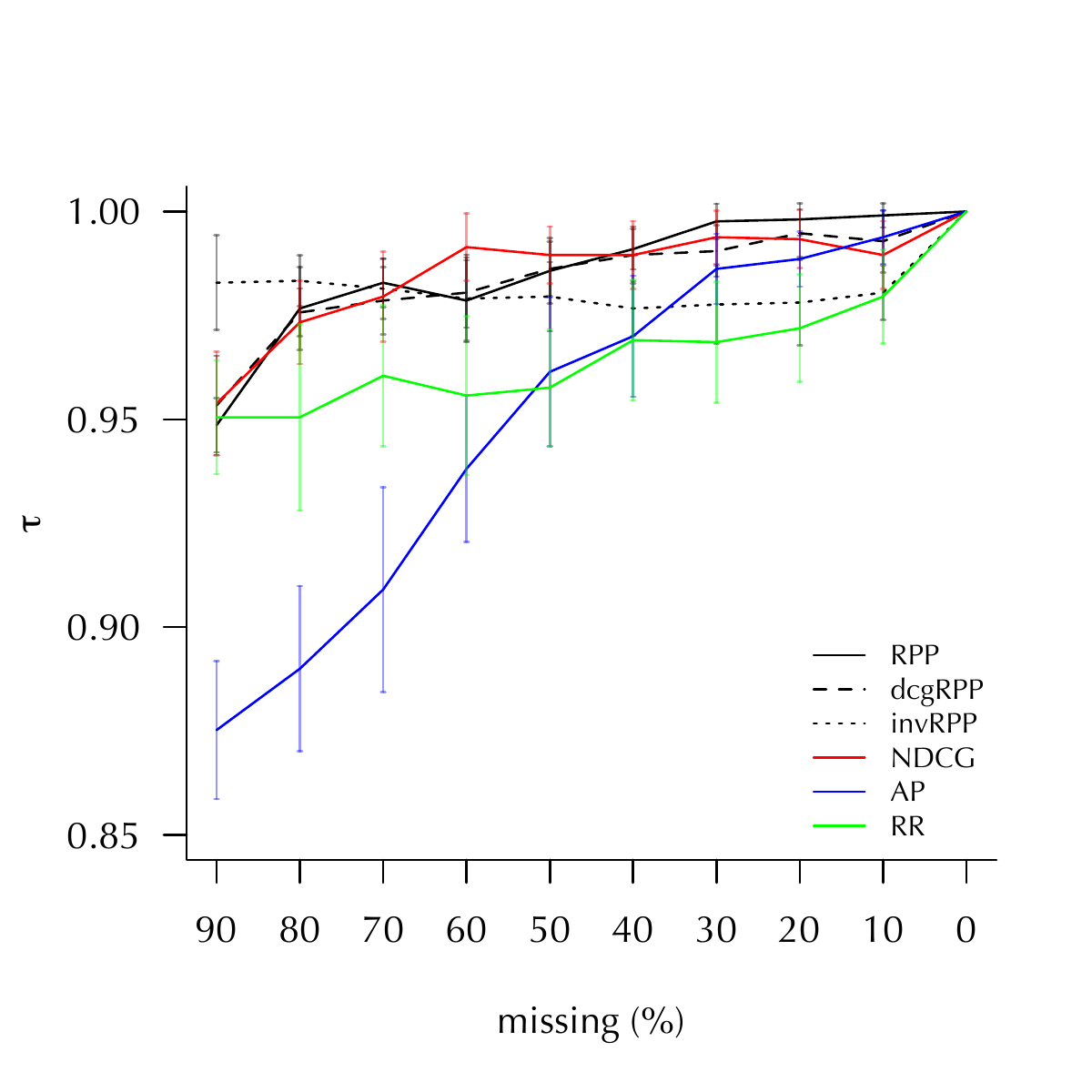}
            \caption{MovieLens 1M}
        \end{subfigure}
    }
    \caption{Kendall's $\tau$ of a system ranking given missing judgments with a system ranking given all judgments. }\label{fig:reduced}
\end{figure}

We show the sensitivity of results when evaluating with fewer labeled items in Figure \ref{fig:reduced}.  Here, for each metric, we calculate the correlation between system rankings with missing labels and system rankings with all labels.  As with missing requests, RR degrades poorly across datasets, which is expected since the removal of the top ranked item is likely to substantially perturb performance (Section \ref{sec:results:metric-correlation}).  AP also degrades poorly, especially when more than 50\% of the judgments are missing.  RPP variants degrade more gracefully and are comparable to NDCG, a metric considered less sensitive to missing label \cite{valcarce:recsys-ranking-metrics-journal}.

\subsection{Discriminative Power}
\label{sec:results:discriminative-power}

We present measurements of discriminative power in Tables \ref{tab:hsd} and \ref{tab:tt}.  Although results are largely consistent for both the HSD test and $t$-test, we include both to further support our analysis.

One fundamental impact we should expect with poorer label efficiency is a reduced ability to distinguish pairs of systems.  Across almost all datasets, we observe that RPP-style preferences have substantially more discriminative power compared to baseline metrics.  Both AP and RR tend to have lower discriminative power than NDCG, consistent with previous results \cite{valcarce:recsys-ranking-metrics-journal}.  The low discriminative power of RR certainly arises from both the poor label efficiency and the large number of ties (Section \ref{sec:results:metric-correlation}).  And, although non-uniform position-weighting (dcgRPP, invRPP) sometimes improves discriminative power slightly, uniform position-weighting (RPP) consistently has high discriminative power compared to baseline metrics.  

The discriminative power of RPP  is present in subtopic evaluations as well, at times dramatically so compared to existing subtopic metrics. We note that this may be, in part, due to the addition of a background pseudo-population reflecting binary relevance.

We observe that the number of detectable differences improves for all methods when more requests are present (e.g. ml-1M, libraryThing, beerAdvocate).  This should be expected since, regardless of metric, more evaluation data will result in better performance estimates.  That said, even in these regimes, RPP-style evaluation is more sensitive.  Moreover, if conducting segment analysis (e.g. for fairness evaluation), under-represented groups, by definition, will have substantially less data.

Although the discriminative power of RPP is not alone sufficient to demonstrate effectiveness, it does provide an important property when considering it for model development or evaluation.  Moreover, given that RPP and its position-weighted variants correlate well with existing metrics, these results suggest that the RPP family may be a more sensitive set of instruments for the same phenomenon.


\section{Discussion}
\label{sec:discussion}
Our experiments were designed to understand if RPP captured properties of existing metrics with the benefit of added sensitivity because of better label efficiency.  Our correlation results (Section \ref{sec:results:metric-correlation}) support the claim that RPP and variants can measure aspects similar to existing metrics while our robustness to incomplete data experiments (Section \ref{sec:results:robustness}) demonstrate that RPP is as robust to incomplete data  as NDCG, an existing metric known to be robust to incomplete data.  Our strongest result suggests that, while capturing these properties of existing metrics, RPP is substantially more sensitive (Section \ref{sec:results:discriminative-power}).  As a result, RPP can complement the existing suite of evaluation metrics, including less sensitive but more realistic metrics based on a domain's user behavior.  

Our philosophy in designing RPP was to minimize the number of assumptions about user behavior, while being flexible enough to model them, as needed.  We demonstrated that incorporating user models through, for example, $p(i)$ could increase correlation with existing metrics (Section \ref{sec:results:metric-correlation}) while maintaining RPP's strong discriminative power (Section \ref{sec:results:discriminative-power}).  We believe that careful incorporation of models of user behavior can further improve the grounding of RPP while preserving its discriminative power.  For example, referring to Figure \ref{fig:metric-vs-preference:preference}, given a labeled preference, we can imagine learning the weights on different pseudo-populations.  Labeled preferences could come from editorial data or from behavioral data such as an interleaving experiment.  The former would be similar to the approach taken by \citet{hassan:machine-assisted-preferences} for top $k$ rankings.  

\citet{chapelle:interleaving} demonstrated the sensitivity of interleaving experiments across a variety of online search scenarios. At the same time, the distribution $p(k)$ is likely to be skewed toward top rank positions, resulting in an under-weighting of higher values of $k$ in Equation \ref{eq:interleaving}.  Because this can result in lower label efficiency, using a more uniform $p(k)$ could improve the sensitivity of online interleaving.

Although we have presented RPP as a way to evaluate systems, how to optimize RPP is an area for future research.  On the one hand, uniformly weighing the importance of different recall levels is similar to methods that train models with a sequence of tasks based on a sampled relevant item  combined with sampled negative items \cite{krichene:sampled-metrics}.

Under these approaches, the model learns to rank all relevant items, weighting them equally.  
In the context of evaluation, this is similar to uniformly weighting all recall levels (i.e. $p(i)=\frac{1}{\nrel}$).  
Our results demonstrate that this may be a more robust way to optimize rankers.

\begin{table}[t]
    \centering
    \caption{Percentage of run differences detected at $p<0.05$ using Tukey's HSD test. }\label{tab:hsd}
    {
        \begin{subtable}[b]{\linewidth}
            \centering
        {\small
            \caption{Single Topic Metrics}\label{tab:hsd:single-topic}
            \begin{tabular}{lccc|ccc}
            &   RPP  & dcgRPP & invRPP & AP    &   NDCG   &   RR  \\
            \hline
            core (2017)	&	\textbf{49.91}	&	49.44	&	40.76	&	36.54	&	39.21	&	14.67	\\
            core (2018)	&	\textbf{52.35}	&	49.22	&	42.33	&	35.29	&	34.66	&	27.97	\\
            deep-docs (2019)	&	\textbf{42.53}	&	40.83	&	30.01	&	33.57	&	41.11	&	6.97	\\
            deep-docs (2020)	&	18.85	&	\textbf{19.35}	&	17.31	&	17.71	&	18.06	&	2.88	\\
            deep-pass (2019)	&	\textbf{42.34}	&	42.19	&	36.34	&	30.03	&	26.73	&	6.76	\\
            deep-pass (2020)	&	45.70	&	\textbf{47.34}	&	45.82	&	30.10	&	20.87	&	22.21	\\
            web (2009)	&	32.00	&	33.69	&	\textbf{35.02}	&	18.35	&	31.56	&	23.23	\\
            web (2010)	&	\textbf{43.75}	&	39.31	&	28.43	&	27.82	&	37.50	&	18.15	\\
            web (2011)	&	\textbf{45.16}	&	43.36	&	34.06	&	31.62	&	40.61	&	14.17	\\
            web (2012)	&	\textbf{41.05}	&	36.44	&	23.49	&	26.06	&	32.09	&	12.77	\\
            web (2013)	&	\textbf{48.42}	&	44.10	&	22.30	&	29.02	&	41.53	&	5.96	\\
            web (2014)	&	48.74	&	48.28	&	36.55	&	44.14	&	\textbf{49.89}	&	18.85	\\
            robust 	&	\textbf{63.47}	&	60.52	&	51.08	&	52.89	&	54.53	&	22.84	\\
            ml-1M&	\textbf{94.29}	&	\textbf{94.29}	&	92.38	&	88.57	&	88.57	&	83.81	\\
            libraryThing	&	96.67	&	97.14	&	\textbf{97.62}	&	95.24	&	95.71	&	93.33	\\            
            beerAdvocate 	&	94.76	&	94.76	&	\textbf{95.71}	&	93.33	&	94.29	&	91.90	\\
            \hline
            \end{tabular}
            }
        \end{subtable}
        
    }
    {\small
        \begin{subtable}[b]{\linewidth}
            \centering
            \caption{Subtopic Metrics}\label{tab:hsd:sub-topic}
            {\small
                \begin{tabular}{lc|ccc}
                &   ST-RPP  & ST-R &   ERR-IA   & MAP-IA    \\
                \hline
                web (2009)	&	\textbf{29.43}	&	28.99	&	24.20	&	17.11	\\
                web (2010)	&	\textbf{40.12}	&	28.02	&	22.18	&	20.77	\\
                web (2011)	&	\textbf{46.11}	&	17.93	&	16.45	&	26.12	\\
                web (2012)	&	\textbf{44.15}	&	12.15	&	16.58	&	25.09	\\
                web (2013)	&	\textbf{49.89}	&	4.48	&	5.52	&	24.04	\\
                web (2014)	&	\textbf{50.11}	&	12.87	&	17.93	&	40.46	\\
                \hline
                \end{tabular}
            }
        \end{subtable}
    }
\end{table}

\begin{table}[t]
    \centering
    \caption{Percentage of run differences detected at $p<0.05$ using the Student's $t$-test with Bonferroni correction. }\label{tab:tt}
    {
        \begin{subtable}[b]{\linewidth}
            \centering
        {\small
            \caption{Single Topic Metrics}\label{tab:tt:single-topic}
            \begin{tabular}{lccc|ccc}
            &   RPP  & dcgRPP & invRPP & AP    &   NDCG   &   RR  \\
            \hline
            core (2017)	&	\textbf{54.13}	&	53.66	&	45.26	&	34.13	&	36.54	&	9.55	\\
            core (2018)	&	\textbf{55.59}	&	53.99	&	48.63	&	35.84	&	40.18	&	21.91	\\
            deep-docs (2019)	&	\textbf{49.64}	&	44.81	&	37.27	&	28.45	&	36.13	&	4.84	\\
            deep-docs (2020)	&	18.01	&	\textbf{19.15}	&	17.36	&	14.34	&	15.58	&	0.50	\\
            deep-pass (2019)	&	43.99	&	\textbf{45.05}	&	42.34	&	24.17	&	23.72	&	1.65	\\
            deep-pass (2020)	&	51.49	&	54.59	&	\textbf{55.93}	&	34.83	&	33.31	&	18.35	\\
            web (2009)	&	34.22	&	\textbf{37.06}	&	36.70	&	15.25	&	34.04	&	19.50	\\
            web (2010)	&	\textbf{52.22}	&	45.16	&	30.04	&	22.58	&	31.25	&	13.71	\\
            web (2011)	&	\textbf{54.42}	&	49.60	&	35.64	&	32.36	&	40.93	&	6.98	\\
            web (2012)	&	\textbf{44.59}	&	39.98	&	24.47	&	22.25	&	32.45	&	10.64	\\
            web (2013)	&	\textbf{59.24}	&	51.26	&	23.01	&	25.52	&	40.82	&	2.51	\\
            web (2014)	&	59.31	&	56.09	&	38.16	&	51.49	&	\textbf{62.30}	&	13.10	\\
            robust 	&	\textbf{65.00}	&	62.29	&	54.21	&	44.82	&	49.34	&	16.21	\\
            ml-1M 	&	\textbf{96.19}	&	95.71	&	\textbf{96.19}	&	91.43	&	94.29	&	85.71	\\
            libraryThing	&	\textbf{98.57}	&	98.10	&	\textbf{98.57}	&	94.29	&	96.67	&	92.38	\\
            beerAdvocate 	&	95.24	&	95.24	&	\textbf{95.71}	&	91.43	&	\textbf{95.71}	&	91.90	\\
            \hline
            \end{tabular}
            }
        \end{subtable}
        
    }
    {\small
        \begin{subtable}[b]{\linewidth}
            \centering
            \caption{Subtopic Metrics}\label{tab:tt:sub-topic}
            {\small
                \begin{tabular}{lc|ccc}
                &   ST-RPP  & ST-R &   ERR-IA   & MAP-IA    \\
                \hline
                web (2009)	&	\textbf{33.24}	&	25.62	&	19.06	&	14.36	\\
                web (2010)	&	\textbf{49.40}	&	16.94	&	14.72	&	11.69	\\
                web (2011)	&	\textbf{56.32}	&	8.99	&	9.52	&	23.64	\\
                web (2012)	&	\textbf{43.71}	&	7.98	&	12.06	&	14.45	\\
                web (2013)	&	\textbf{60.77}	&	2.35	&	2.90	&	19.45	\\
                web (2014)	&	\textbf{63.68}	&	9.89	&	12.18	&	46.44	\\
                \hline
                \end{tabular}
            }
        \end{subtable}
    }
\end{table}

\section{Conclusion}
\label{sec:conclusion}
We have presented recall-paired preference (RPP), a method for evaluating rankings that avoids first computing an evaluation metric.  Through extensive experimentation, we demonstrate that RPP reflects many properties of existing metrics with a substantially improved sensitivity.  We believe that this approach can be extended in multiple directions, including refining user models while preserving its sensitivity.

\bibliographystyle{ACM-Reference-Format}
\bibliography{rpp.bib}
\end{document}